\documentclass[a4paper,11pt]{article}

\usepackage[numbers,sort&compress]{natbib}
\usepackage[all]{xy}
\usepackage{amsmath,amsfonts,amssymb,amsthm,epsfig,amscd,comment,latexsym,psfrag}
\usepackage{CJK}
\usepackage{mathrsfs}
\usepackage{nicefrac,xspace,tikz}
\usepackage{arydshln}
\numberwithin{equation}{section}
\usetikzlibrary{arrows}
\usepackage{graphicx}
\usepackage{caption,subcaption}
\usepackage{xcolor,float}
\usepackage{appendix}
\usepackage{graphicx}
\usepackage{caption}
\makeatletter
\newcommand{\rmnum}[1]{\romannumeral #1}
\newcommand{\Rmnum}[1]{\expandafter\@slowromancap\romannumeral #1@}
\makeatother

\def\footnoterule{\kern 1mm \hrule width 7cm \kern 2.2mm}%

\def\dprod{\displaystyle\prod}
\def\dlim{\displaystyle\lim}
\def\dsum{\displaystyle\sum}

\def\Tr{\mathrm{Tr}}

\renewcommand{\b}{\beta}

\allowdisplaybreaks
\begin{document}
\begin{titlepage}
\begin{center}
{\Large\bf $W$-representation of Rainbow tensor model}\vskip .2in
{\large Bei Kang$^{a,}$\footnote{kangbei@ncwu.edu.cn}
Lu-Yao Wang$^{b,}$\footnote{wangly100@outlook.com}  Ke Wu$^{b,}$\footnote{wuke@cnu.edu.cn},
Jie Yang $^{b,}$\footnote{yangjie@cnu.edu.cn},
Wei-Zhong Zhao$^{b,}$\footnote{Corresponding author: zhaowz@cnu.edu.cn}} \vskip .2in
$^a${\em School of Mathematics and Statistics, North China University of Water Resources and Electric Power,
Zhengzhou 450046, Henan, China}\\
$^b${\em School of Mathematical Sciences, Capital Normal University,
Beijing 100048, China} \\

\begin{abstract}
We analyze the rainbow tensor model and present the Virasoro constraints,
where the constraint operators obey the Witt algebra and null 3-algebra.
We generalize the method of $W$-representation in matrix model to the rainbow tensor model,
where the operators preserving and increasing the grading play a crucial role.
It is shown that the rainbow tensor model can be realized by acting on
elementary function with exponent of the operator increasing the grading.
We derive the compact expression of correlators and apply it to several models,
i.e., the red tensor model, Aristotelian tensor model and $r=4$ rainbow tensor
model. Furthermore, we discuss the case of the non-Gaussian red tensor model
and present a dual expression for partition function through differentiation.

\end{abstract}

\end{center}


\end{titlepage}

\section{Introduction}

$W$-representation of matrix model which realizes partition function by acting on
elementary functions with exponents of the given $W$-operator has attracted considerable attention.
It indeed gives a dual expression for partition function
through differentiation rather than integration.
As the fundamental matrix models, it turned out that
the Gaussian Hermitian and complex matrix
models can be written as the form of the $W$-representations \cite{Shakirov2009}-\cite{Itoyama2020}.
Since these $W$-representations can be expressed in terms of characters,
the corresponding matrix models are reformulated as the sum of Schur functions over all Young diagrams.
For the $\beta$-deformed Gaussian Hermitian and complex models, their $W$-representations still exist.
The character expansions of these models can be given by the Jack polynomials.
The studies of $W$-representations  have also been devoted to the supersymmetric generalizations of
matrix models, i.e., supereigenvalue models \cite{Chen,wang2020}.

As the generalizations of matrix models from matrices to tensor,
tensor models become very useful in the deep study of higher dimensional quantum gravity
\cite{Jonsson}-\cite{Sasakura}. Quite recently, the operators/Feynman diagrams
correspondence in quantum field theory was provided \cite{Amburg}.
For the number of Feynman diagrams with $n$ propagators in the rank $r-1$
complex tensor model, it is equal to the number of singlet operators with $2n$ vertices
in the rank $r$ complex tensor model.
Tensor models are also very interesting in their own right \cite{Gurau}-\cite{ItoyamaPLB2020}.
Recently the tensorial generalization of characters \cite{ItoyamaPLB2019,ItoyamaJHEP2019}
and correlators in tensor models from character calculus \cite{MironovPLB2017}-\cite{ItoyamaPLB2020}
have been analyzed.
The Gaussian tensor model is a model of complex $r$-tensors with the Gaussian action.
It can be expressed as the forms of the characters and $W$-representation \cite{ItoyamaPLB2020},
\begin{eqnarray}\label{partitionfunctionr}
Z_r\{p^{(i)}\}&=&\sum_{R_1,\cdots,R_r}\sum_{\bigtriangleup\vdash n}
\frac{\prod_{i=1}^r\psi_{R_i}(\Delta)}{z_{\Delta}}\prod_{m=1}^r[\chi_{R_m}\{p^{(m)}\}\cdot\prod_{(i,j)\in R_m}(N_m+i-j)]\nonumber\\
&=&e^{\hat{\mathcal{W}}(N_1,\cdots,N_r)}\cdot 1,
\end{eqnarray}
where $\psi_R(\Delta)$, $z_{\Delta}$ and $\chi_{R_m}\{p^{(m)}\}$
are respectively the character of symmetric group, the symmetry factor of Young diagram $\Delta$
and the Schur function as a function of time-varibles $p_k$,
the operator $\hat{\mathcal{W}}(N_1,\cdots,N_r)$ is given by
\begin{eqnarray}
\hat{\mathcal{W}}(N_1,\cdots,N_r)&=&\hat{O}_1(N_1)\cdots\hat{O}_m(N_m)
\circ\sum_{k}\frac{\prod_{m=1}^{r}p_k^{(m)}}{k}\circ\hat{O}_1^{-1}(N_1)\cdots\hat{O}^{-1}_m(N_m),
\end{eqnarray}
here the subscript $m$ of $\hat{O}_m(N)$ means that this operator acts on the variables $p_k^{(m)}$,
the operator $\hat{O}(N)$ satisfies $\hat{O}(N)\chi_R=\frac{D_R(N)}{d_R}\chi_R$,
$D_R(N)=\chi_R\{p_k=N\}$ is the dimension of the linear group and $d_R=\chi_R\{p_k=\delta_{1,k}\}$.
The generalized characters which generate the partition function~(\ref{partitionfunctionr})
form an over-complete basis in the space of all gauge invariant operators with non-vanishing
Gaussian averages.
It should be noted that in the generic tensor model, there is no simple way to remove the redundancy.

The Aristotelian rainbow tensor model with a single complex tensor of rank 3 and the RGB (red-green-blue)
symmetry is the simplest of the rainbow tensor models \cite{ItoyamaJHEP2017}-\cite{ItoyamaNPB2018}.
Recently, with the example of the Aristotelian tensor model,
Itoyama et al. \cite{ItoyamaJHEP2017}
introduced a few methods which allow one to connect calculations in the tensor models to those
in the matrix models.
Well known is that the partition functions of various matrix models can be realized by
the $W$-representations, where the operators preserving and increasing
the grading play a crucial role \cite{Shakirov2009}.
The goal of this paper is to make a step towards the $W$-representation
of the rainbow tensor model. We present its $W$-representation and
give the compact expression of correlators.

This paper is organized as follows. In section 2, we show that the rainbow tensor model
can be realized by acting on elementary function with exponent of the given operator.
The Virasoro constraints are also presented. Then we derive the compact expressions of the correlators.
In sections 3 and 4, we focus on the correlators in the Aristotelian and $r=4$ tensor models, respectively.
In section 5, we consider the (Non-Gaussian) red tensor model. We end this
paper with the conclusions in section 6.

\section{$W$-representation of rainbow tensor model}

For the rainbow model with the rank~$r$ complex tensors and with the gauge symmetry~
$\mathcal{U}=U(N_1)\otimes\cdots\otimes U(N_r)$,
the gauge-invariant operators of level~$n$ are given by \cite{ItoyamaNPB2018}
\begin{eqnarray}\label{ksigma}
\mathcal{K}_{\sigma}^{(n)}=
\mathcal{K}_{\sigma_1\otimes\cdots\otimes\sigma_n}^{(n)}= \prod_{p=1}^{n}A_{i^{(p)}}^{j^{(p)}_1,\cdots,j^{(p)}_{r-1}}
\bar A_{j^{\sigma_2(p)}_1,
\cdots,j^{\sigma_r(p)}_{r-1}}^{i^{\sigma_1(p)}},
\end{eqnarray}
where $A_i^{j_1,\ldots,j_{r-1}}$ is a tensor of rank~$r$ with one covariant and~$r-1$
contravariant indices, its conjugate tensor is $\bar{A}^i_{j_1,\ldots,j_{r-1}}$,
$\sigma$ is an element of the double coset $\mathcal{S}_n^r= S_n\backslash S_n^{\otimes r}/S_n$ and $\deg\sigma=n$.
Here the different types of indices in the fields and fields themselves are assigned with different color.
We may choose some operators in (\ref{ksigma}) to generate a graded ring of gauge
invariant operators  with addition, multiplication, cut and join operations. These operators are called keystones.
The connected operators in this ring can generate the renormalization group (RG) completed rainbow tensor model \cite{ItoyamaNPB2018}.

Let us introduce the RG-completed rainbow tensor model
\begin{eqnarray}\label{rainbowrg}
Z_{R}&=&
\int dA d\bar A \exp(-\mu \Tr A\bar A+\sum_{n=1}^{\infty}\sum_{\deg\sigma=n}t_{\sigma}^{(n)}\mathcal{K}_{\sigma}^{(n)})\nonumber\\
&=&\sum_{s=0}^{\infty}Z_{R}^{(s)},
\end{eqnarray}
where $\mu$ is a constant, $t_{\sigma}^{(n)}$ are the time variables,
the measure is induced by the norm $\parallel \delta A \parallel^2=\delta A_i^{j_1,\ldots,j_{r-1}} \delta \bar{A}^i_{j_1,\ldots,j_{r-1}}$,
\begin{eqnarray}
Z_{R}^{(s)}=\int dA d\bar A \exp(-\mu \Tr A\bar A)\cdot\sum_{l=0}^{\infty}\sum_{n_1+\cdots +n_l=s}
\frac{1}{l!}\langle\mathcal{K}_{\sigma_1}^{(n_1)}\mathcal{K}_{\sigma_2}^{(n_2)}\cdots\mathcal{K}_{\sigma_l}^{(n_l)}
\rangle t_{\sigma_1}^{(n_1)}t_{\sigma_2}^{(n_2)}\cdots t_{\sigma_l}^{(n_l)},
\end{eqnarray}
and the correlators~$\langle\mathcal{K}_{\sigma_1}^{(n_1)}\mathcal{K}_{\sigma_2}^{(n_2)}\cdots\mathcal{K}_{\sigma_l}^{(n_l)}
\rangle $ are defined by
\begin{eqnarray}
\langle\mathcal{K}_{\sigma_1}^{(n_1)}\mathcal{K}_{\sigma_2}^{(n_2)}\cdots\mathcal{K}_{\sigma_l}^{(n_l)}
\rangle =\frac{\int dA d\bar A \mathcal{K}_{\sigma_1}^{(n_1)}\mathcal{K}_{\sigma_2}^{(n_2)}
\cdots\mathcal{K}_{\sigma_l}^{(n_l)} \exp(-\mu \Tr A\bar A)}{\int dA d\bar A  \exp(-\mu \Tr A\bar A)}.
\end{eqnarray}

For any connected operator $\mathcal{K}_{\alpha}^{(a)}$  in the exponent of~(\ref{rainbowrg}),
let us consider the deformation
$\delta A=\dsum_{a=1}^{\infty}\dsum_{deg\alpha=a}t_{\alpha}^{(a)}\dfrac{\partial
\mathcal{K}_{\alpha}^{(a)}}{\partial \bar A}$ of the integration variable in the integral (\ref{rainbowrg}).
It gives
\begin{eqnarray}\label{ward}
&&\int dA d\bar A [\dsum_{a=1}^{\infty}\dsum_{\deg\alpha=a}t_{\alpha}^{(a)}\Delta \mathcal{K}_{\alpha}^{(a)}
+\dsum_{a,n=1}^{\infty}\dsum_{\deg\alpha=a}\sum_{\deg\sigma=n}t_{\alpha}^{(a)}t_{\sigma}^{(n)}\{\mathcal{K}_{\sigma}^{(n)},\mathcal{K}_{\alpha}^{(a)}\}\nonumber\\
&&-\mu\dsum_{a=1}^{\infty}\dsum_{\deg\alpha=a}at_{\alpha}^{(a)} \mathcal{K}_{\alpha}^{(a)}]
\exp(-\mu \Tr A\bar A+\dsum_{n=1}^{\infty}\sum_{\deg\sigma=n}t_{\sigma}^{(n)}\mathcal{K}_{\sigma}^{(n)})=0,
\end{eqnarray}
where $\Delta$ and $\{\}$ are respectively the cut and join operations,
the actions of the cut and join operations on the gauge-invariant operators are
\begin{eqnarray}\label{cut}
\Delta \mathcal{K}_{\alpha}^{(a)}&=&\sum_{i=1}^{N_1}
\sum_{j_1=1}^{N_2}\cdots
\sum_{j_{r-1}=1}^{N_{r}}
\dfrac{\partial^2 \mathcal{K}_{\alpha}^{(a)}}{\partial A_i^{j_1,\ldots,j_{r-1}}
\partial \bar{A}^i_{j_1,\ldots,j_{r-1}}}\nonumber\\
&=&\sum_{k=1}^r\sum_{\substack{\beta_1,\cdots,\beta_k\\ b_1+\cdots+b_k+1=a}}\Delta_{\alpha}^{\beta_1,\cdots,\beta_k}
\mathcal{K}_{\beta_1}^{(b_1)}\cdots\mathcal{K}_{\beta_k}^{(b_k)},\ a\geqslant 2,
\end{eqnarray}
and
\begin{eqnarray}\label{join}
\{\mathcal{K}_{\sigma}^{(n)},\mathcal{K}_{\alpha}^{(a)}\}&=&
\sum_{i=1}^{N_1}
\sum_{j_1=1}^{N_2}\cdots
\sum_{j_{r-1}=1}^{N_{r}}
\dfrac{\partial \mathcal{K}_{\sigma}^{(n)}}{\partial A_i^{j_1,\ldots,j_{r-1}}}
\dfrac{\partial \mathcal{K}_{\alpha}^{(a)}}{ \partial \bar{A}^i_{j_1,\ldots,j_{r-1}}}\nonumber\\
&=&
\sum_{\deg\beta=n+a-1}
\gamma_{\sigma,\alpha}^{\beta}
\mathcal{K}_{\beta}^{(n+a-1)},
\end{eqnarray}
$\Delta_{\alpha}^{\beta_1,\cdots,\beta_k}$ and $\gamma_{\sigma,\alpha}^{\beta}$ are
the coefficients.

From (\ref{ward}), we may deduce that the partition function (\ref{rainbowrg}) satisfies
\begin{eqnarray}\label{dw}
\mu \hat{D}_{r}Z_{R}=\hat{W}_rZ_{R},
\end{eqnarray}
where
\begin{eqnarray}\label{operatorD}
\hat{D}_{r}&=& \dsum_{a=1}^{\infty}\dsum_{\deg\alpha=a}at_{\alpha}^{(a)}\frac{\partial}{\partial t_{\alpha}^{(a)}},
\end{eqnarray}
\begin{eqnarray}\label{operator}
\hat{W}_r&=&\dsum_{a,n=1}^{\infty}\dsum_{\deg\alpha=a}\sum_{\deg\sigma=n}\sum_{\deg\beta=n+a-1}\gamma_{\sigma,\alpha}^{\beta}
 t_{\alpha}^{(a)} t_{\sigma}^{(n)}\frac{\partial}{\partial t_{\beta}^{(n+a-1)}}+t_{\underbrace{id\otimes \cdots\otimes id}_r}^{(1)}N_1\cdots N_r\nonumber\\
&&+\dsum_{a=1}^{\infty}\dsum_{\deg\alpha=a}\sum_{k=1}^r\sum_{\substack{\beta_1,\cdots,\beta_k\\b_1+\cdots+b_k+1=a}}(1-\delta_{a,1})
\Delta_{\alpha}^{\beta_1,\cdots,\beta_k}t_{\alpha}^{(a)}
\frac{\partial}{\partial t_{\beta_1}^{(b_1)}}\cdots\frac{\partial}{\partial t_{\beta_k}^{(b_k)}}.
\end{eqnarray}
The commutation relation between~$\hat{D}_r$ and~$\hat{W}_r$ is
\begin{eqnarray}\label{commu}
[\hat{D}_{r},\hat{W}_r]=\hat{W}_r.
\end{eqnarray}
In terms of the operators $\hat{D}_{r}$ and $\hat{W}_{r}$, we may introduce the Virasoro constraints
\begin{eqnarray}\label{Vcons}
L_{m}Z_{R}=0,
\end{eqnarray}
where the constraint operators $L_{m}$ are given by
\begin{eqnarray}
L_m=-\frac{1}{\mu}\hat{W}_r^m(\hat{W}_r-\mu\hat{D}_{r}), \ m\in \mathbb{N},
\end{eqnarray}
which yield the Witt algebra
\begin{eqnarray}\label{Valg}
[L_m, L_n]=(n-m)L_{m+n},
\end{eqnarray}
and null 3-algebra
\begin{eqnarray}\label{3Valg}
[L_k, L_m, L_n]=0.
\end{eqnarray}

Let us consider the operators $\hat D_{r}$ and $\hat{W}_{r}$ acting on $Z_R^{(s)}$, respectively.
We have
\begin{eqnarray}\label{dzs}
\hat {D}_{r} Z_R^{(s)}=sZ_R^{(s)},
\end{eqnarray}
\begin{eqnarray}\label{increa}
\hat{W}_rZ_R^{(s)}=\mu(s+1)Z_R^{(s+1)}.
\end{eqnarray}
It is similar with the case of the Gaussian hermitian model \cite{Shakirov2009}.
We immediately recognize that the operators $\hat D_{r}$ and $\hat{W}_{r}$
are indeed the operators preserving and increasing the grading, respectively.
Thus the partition function can be realized by acting on elementary function with exponents of the operator $\hat{W}_r$
\begin{eqnarray}\label{exp}
Z_R=\exp(\frac{1}{\mu }\hat{W}_r)\cdot 1.
\end{eqnarray}

As done in the matrix models, we formally write
the $m$-th power of the operator $\hat{W}_r$ as
\begin{eqnarray}\label{wm}
\hat{W}_{r}^m&=&\sum_{i=1}^{2m}\sum_{j=1}^{rm}\sum_{\substack{a_1+\cdots+a_i=\\b_1+\cdots+b_j+m}}
(P_r)^{\alpha_1,\cdots,\alpha_i} _{ \beta_1,\cdots,\beta_j}\sum_{\deg\alpha_i=a_i}\sum_{\deg\beta_j=b_j} t_{\alpha_1}^{(a_1)}\cdots t_{\alpha_{i}}^{(a_i)}
\frac{\partial}{\partial {t_{\beta_1}^{(b_1)}}}
\cdots \frac{\partial}{\partial {t_{\beta_{j}}^{(b_j)}}}\nonumber\\
&&
+\sum_{i=1}^{m}\sum_{a_1+\cdots+a_i=m}\sum_{\deg\alpha_i=a_i}
P_{r}^{\alpha_1,\cdots,\alpha_i} t_{\alpha_1}^{(a_1)}\cdots t_{\alpha_{i}}^{(a_i)},
\end{eqnarray}
where the coefficients  $P_{r}^{\alpha_1,\cdots,\alpha_i}$  and $(P_r)^{\alpha_1,\cdots,\alpha_i} _{ \beta_1,\cdots,\beta_j}$ are the polynomials of ~$N_1,\cdots,N_r$.

Substituting (\ref{wm}) into (\ref{exp}) and comparing the coefficients of~$t_{\alpha_1}^{(a_1)}\cdots t_{\alpha_{i}}^{(a_i)}$ in the expansion of (\ref{exp})
with the corresponding terms in (\ref{rainbowrg}), we finally derive the compact expression of correlators
\begin{eqnarray}\label{corrf}
\left\langle \mathcal{K}_{\alpha_1}^{(a_1)}\cdots \mathcal{K}_{\alpha_{i}}^{(a_i)} \right\rangle=\frac{i!}{\mu ^mm!\lambda_{(\alpha_1,\cdots,\alpha_i)}}\sum_{\tau}
P_{r}^{\tau(\alpha_1),\cdots,\tau(\alpha_i)},
\end{eqnarray}
where~$m=a_1+\cdots+a_i$, $\tau$ denotes all distinct permutations of $(\alpha_1,\cdots,\alpha_i)$ and~$\lambda_{(\alpha_1,\cdots,\alpha_i)}$
is the number of $\tau$ with respect to~$\alpha_1,\cdots,\alpha_i$.

Let us turn to the Virasoro constraints (\ref{Vcons}). It can be rewritten as
\begin{eqnarray}\label{wlz}
\hat{W}_r^{m} Z_R=\mu^{m}\prod_{j=0}^{m-1}(\hat{D}_{r}-j) Z_R, \ m\in \mathbb{N^*}.
\end{eqnarray}
Since the coefficients
of~$t_{\alpha_1}^{(a_1)}\cdots t_{\alpha_{i}}^{(a_i)}$ on both sides in (\ref{wlz}) with $\sum_{j=1}^{i}a_{j}=m$
are equal, we can not only derive the correlators (\ref{corrf}), but also
the exact correlators
\begin{eqnarray}\label{spcorrf}
\langle (\mathcal{K}_{1} )^{i} \rangle=\frac{1}{\mu^i}P_{r}^{\overbrace{{1},\cdots,{1}}^i}
=\frac{1}{\mu^i}\prod_{j=0}^{i-1}(\mathcal{N}_r+j),
\end{eqnarray}
where
\begin{eqnarray}
P_{r}^{\overbrace{{1},\cdots,{1}}^i}=
(i-1+\mathcal{N}_r)P_{r}^{\overbrace{{1},\cdots,{1}}^{i-1}}=
\cdots=\prod_{j=0}^{i-1}(\mathcal{N}_r+j),
\end{eqnarray}
$\mathcal{N}_r=\dprod_{i=1}^rN_{i}$, and $\mathcal{K}_1=A_i^{j_1,\cdots,j_{r-1}}\bar{A}_{j_1,\cdots,j_{r-1}}^i$.

For the Gaussian average of the rank $r$ operator $O^{(r)}$ in the
rank $r$ complex tensor model, there is a limit relation with $\langle O^{(r+1)}\rangle_{r+1}$
in the rank $r+1$ model \cite{Amburg}, i.e.,
%$<O^{(r-1)}>=\dlim_{N_i\rightarrow\infty}\dfrac{1}{N_i^n}\sum_{ops}<O^{(r)}>$ \cite{Amburg}, where $\dsum_{ops}$ denotes the sum over the operators $O^{(r)}$.???
%For the case of the correlators, we have
\begin{eqnarray}
\left\langle \mathcal{K}_{\alpha_1}^{(a_1)}\cdots \mathcal{K}_{\alpha_{i}}^{(a_i)} \right\rangle_r
=
\dlim_{N_{r+1}\rightarrow\infty}\frac{1}{N_{r+1}^m}
\sum_{\sigma_{r+1}\in S_m}
%\sum_{\substack{\beta_1\oplus\cdots\oplus\beta_j
%=\\ \alpha_1\oplus\cdots\oplus\alpha_i\oplus\sigma_{r+1}}}
\left\langle \prod_{p=1}^m W_{p,\sigma_{r+1}(p)}(\mathcal{K}_{\alpha_1}^{(a_1)}\cdots \mathcal{K}_{\alpha_{i}}^{(a_i)}) \right\rangle_{r+1},
\end{eqnarray}
 where $a_1+\cdots+a_i=m$, $S_m$ is the symmetric group that consists of permutations of $m$ elements,
 $W_{p,q}(A^{(p)}\bar{A}_{(q)})=A_{a_1,\cdots,a_r}^{(p)}\bar{A}_{(q)}^{b_1,\cdots,b_r}
 =\prod_{i=1}^r\delta_{a_i}^{b_i}$ is the Wick contractions of the $p$-th $A$ and $q$-th $\bar{A}$
in $\mathcal{K}_{\alpha_1}^{(a_1)}\cdots \mathcal{K}_{\alpha_{i}}^{(a_i)}$.

When particularized to the special correlators $\langle (\mathcal{K}_{1} )^{i} \rangle_r$, we have
\begin{eqnarray}\label{pcr}
\dlim_{N_{r+1}\rightarrow\infty}\frac{1}{N_{r+1}^i}
\sum_{\sigma_{r+1}\in S_i}
%\sum_{\substack{\beta_1\oplus\cdots\oplus\beta_j=\\ \underbrace{id\oplus\cdots\oplus id}_i\oplus\sigma_{r+1}}}
\left\langle
\prod_{p=1}^i W_{p,\sigma_{r+1}(p)}(\underbrace{\mathcal{K}_{id}^{(1)}\cdots\mathcal{K}_{id}^{(1)}}_{i})
\right\rangle_{r+1}
=\frac{1}{\mu^i}\prod_{j=0}^{i-1}(\mathcal{N}_r+j).
\end{eqnarray}

Taking $i=1,2$ and $3$ in (\ref{pcr}), respectively, it gives
\begin{eqnarray*}
\dlim_{N_{r+1}\rightarrow\infty}\frac{1}{N_{r+1}}\mathcal{N}_{r+1}&=&\mathcal{N}_r,
\end{eqnarray*}
\begin{eqnarray*}
\dlim_{N_{r+1}\rightarrow\infty}\frac{1}{N_{r+1}^2}(P_{r+1}^{\overbrace{id\otimes\cdots \otimes id}^{r} \otimes(12)})=2\mathcal{N}_r,
\end{eqnarray*}
\begin{eqnarray}
&&\dlim_{N_{r+1}\rightarrow\infty}\frac{1}{N_{r+1}^3}[
2P_{r+1}^{\overbrace{id\otimes\cdots \otimes id}^{r}\otimes (123)}+
3P_{r+1}^{id,\overbrace{id\otimes\cdots \otimes id}^{r} \otimes (12)}\nonumber\\
&&+3P^{\overbrace{id\otimes\cdots \otimes id}^{r} \otimes (12),id}_{r+1}]
=18\mathcal{N}_r^2+12\mathcal{N}_r.
\end{eqnarray}

\section{Correlators in the Aristotelian tensor model}

In the Aristotelian model with the tensor~$A_{\textcolor{red}{i}}^{{\textcolor{green}{j_1}},{\textcolor{blue}{j_2}}}$ of rank~$r=3$,
the ring is generated by keystone operators \cite{ItoyamaJHEP2017}
\begin{eqnarray*}
\includegraphics[height=1.9cm]{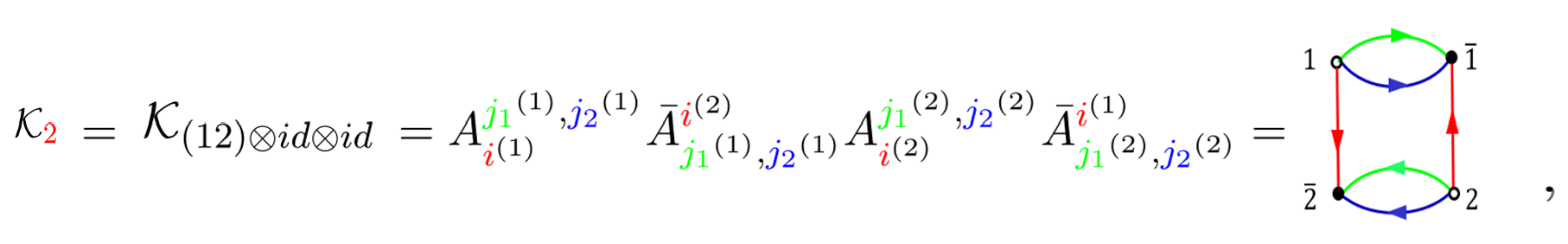}
\end{eqnarray*}
\begin{eqnarray}
\includegraphics[height=2cm]{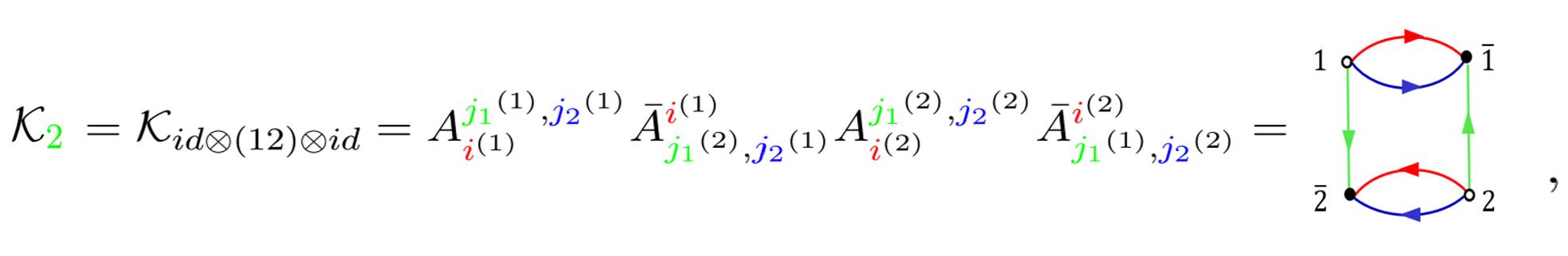}
\end{eqnarray}
where $1$ and $\bar{1}$ represent the first two fields $A$ and $\bar{A}$, $2$ and $\bar{2}$
represent the last two fields $A$ and $\bar{A}$, respectively,
the vertices are fields (tensors), the different color thin lines represent the contraction of indices in the operators,
and the directions of arrows depend on the choice of covariant and contravariant indices.
Note that the ring contains the tree and loop operators.
If the operator belongs to the sub-ring generated only by the
join operation (\ref{join}), this operator is a tree operator, otherwise, it is a loop operator.

\begin{figure}[H]
\centering
\begin{minipage}[c]{0.4\textwidth}
\centering
\includegraphics[height=1.9cm]{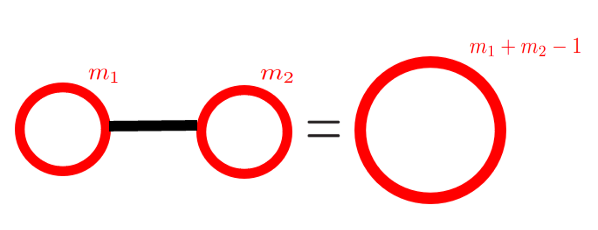}
\caption*{(a)}
\end{minipage}
\begin{minipage}[c]{0.4\textwidth}
\centering
\includegraphics[height=1.8cm]{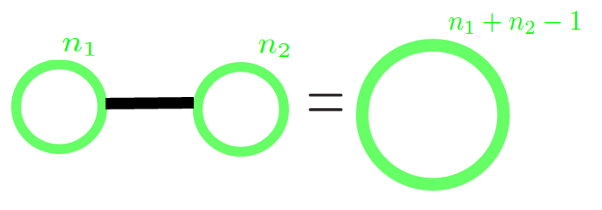}
\caption*{(b)}
\end{minipage}
\nonumber\\
\begin{minipage}[c]{0.4\textwidth}
\centering
\includegraphics[height=1.9cm]{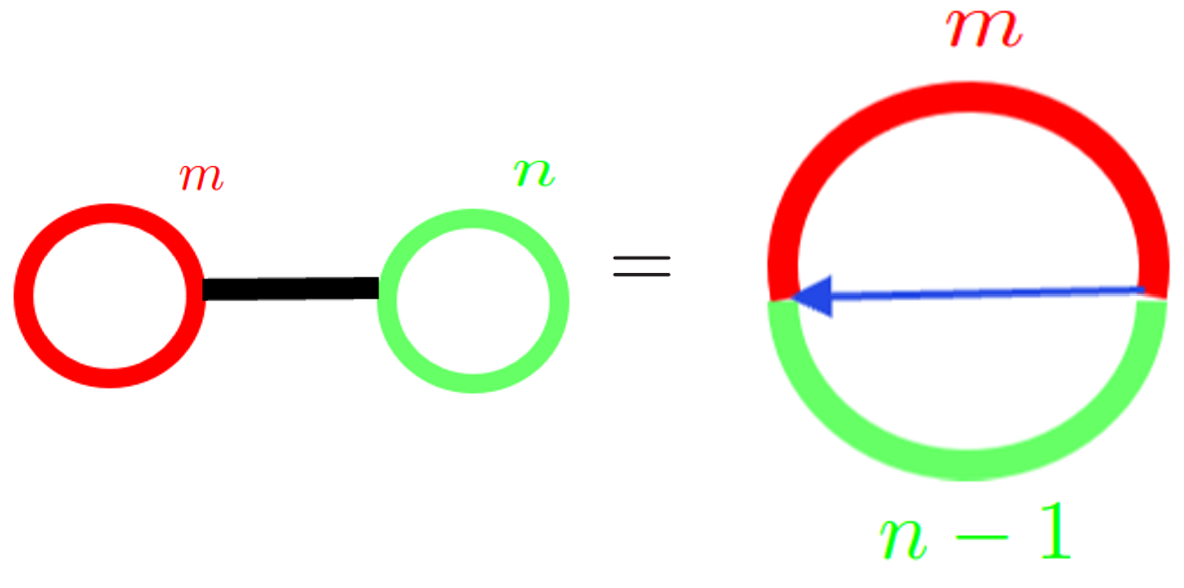}
\caption*{(c)}
\end{minipage}
\begin{minipage}[c]{0.4\textwidth}
\centering
\includegraphics[height=1.9cm]{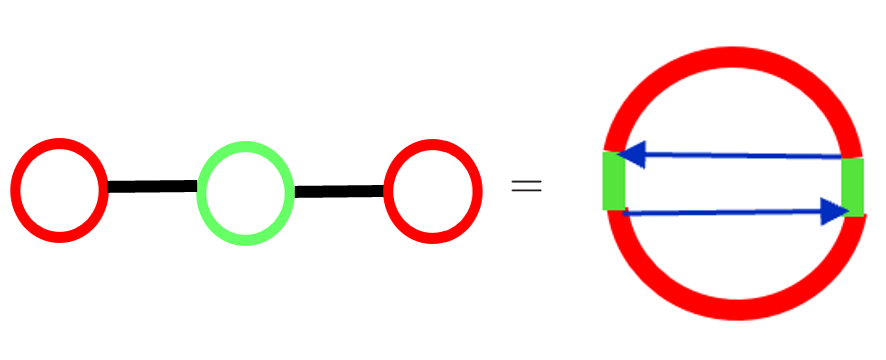}
\caption*{(d)}
\end{minipage}
\caption{Diagrams of tree operators from Ref.\cite{ItoyamaJHEP2017}.}
\label{alltree}
\end{figure}

The tree operators made from $\mathcal{K}_{\textcolor{red}{2}}$ or $\mathcal{K}_{\textcolor{green}{2}}$ alone are constructed by merging two vertices
in two thick cirles (propagators) of the same color (figures.\ref{alltree} (a), (b)).
When tree operators involve chains with both $\mathcal{K}_{\textcolor{red}{2}}$ and $\mathcal{K}_{\textcolor{green}{2}}$,
they are constructed by merging two vertices of two thick circles (propagators) of different colors, two tree operators are drawn in figures.\ref{alltree} (c) and (d).
It is known that all tree-operators are single planar cycles, and these operators which are depicted as one connected diagram are called the connected operators.

\begin{figure}[H]
\centering
\begin{minipage}[c]{0.4\textwidth}
\centering
\includegraphics[height=2.6cm]{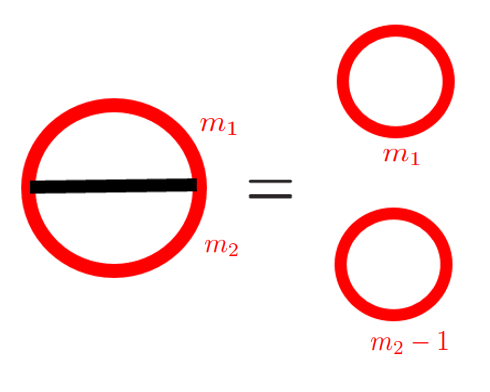}
\caption*{(a)}
\end{minipage}
\begin{minipage}[c]{0.4\textwidth}
\centering
\includegraphics[height=2.6cm]{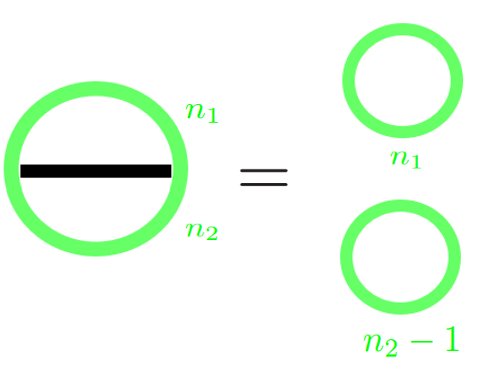}
\caption*{(b)}
\end{minipage}
\nonumber\\
\begin{minipage}[c]{0.4\textwidth}
\centering
\includegraphics[height=1.5cm]{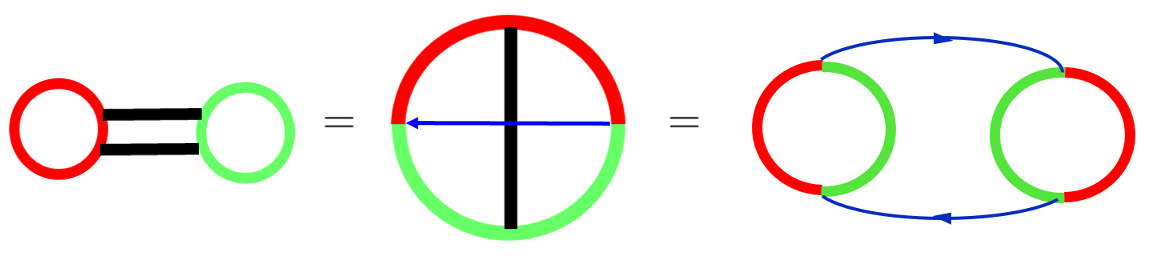}
%${\textcolor{red}{\mathcal{K}_n}}$}
%\label{tu7}
\caption*{(c)}
\end{minipage}%
\begin{minipage}[c]{0.4\textwidth}
\centering
\includegraphics[height=2cm]{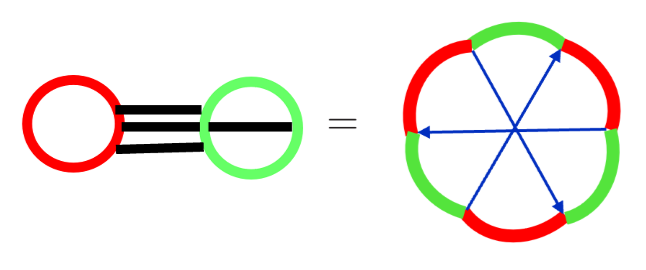}
\caption*{(d)}
\end{minipage}
\caption{Diagrams of loop operators from Ref.\cite{ItoyamaJHEP2017}.}
\label{loopr=3}
\end{figure}

The loop operators made from $\mathcal{K}_{\textcolor{red}{2}}$ or $\mathcal{K}_{\textcolor{green}{2}}$ alone are constructed by merging two vertices
inside a thick circle (propagator). Two such loop operators are drawn in figures.\ref{loopr=3} (a) and (b),
they are disconnected collections of red or green circles. These operators which are depicted as disconnected collection of some diagrams are called disconnected operators.
When the loop operators involve both $\mathcal{K}_{\textcolor{red}{2}}$ and $\mathcal{K}_{\textcolor{green}{2}}$, they are either the red-green cycles
with the intersecting blue shortcuts or several such red-green cycles with the shortcuts connected by thin blue lines.
Two loop operators are drawn in figures.\ref{loopr=3} (c) and (d),  which are constructed by merging two vertices in two thick circles (propagators) of two different colors.

Here the thick black line represents the Feynman propagator,
$\mathcal{K}_{\textcolor{red}{m}}=\mathcal{K}_{(12\cdots m)\otimes id\otimes id}$ and $\mathcal{K}_{\textcolor{green}{m}}=\mathcal{K}_{id\otimes(12\cdots m)\otimes id}$
are depicted as the red and green circles of length $m$
\begin{eqnarray*}
\centering
\includegraphics[height=3cm]{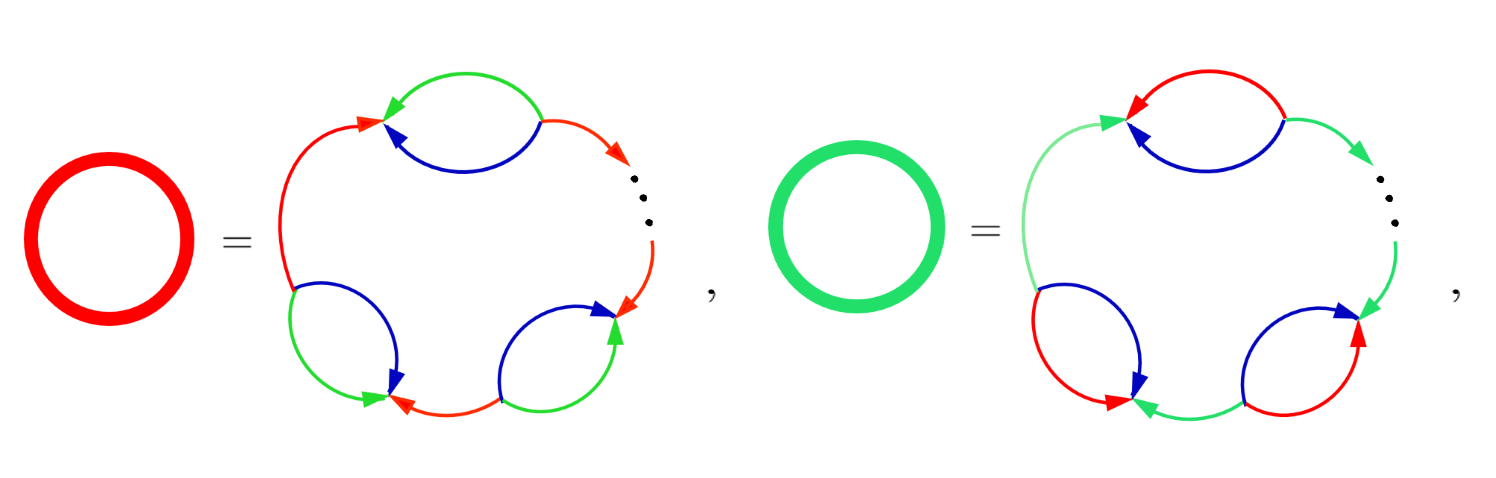}
\end{eqnarray*}
and the thick red and green lines are
\begin{eqnarray*}
\centering
\includegraphics[height=2.45cm]{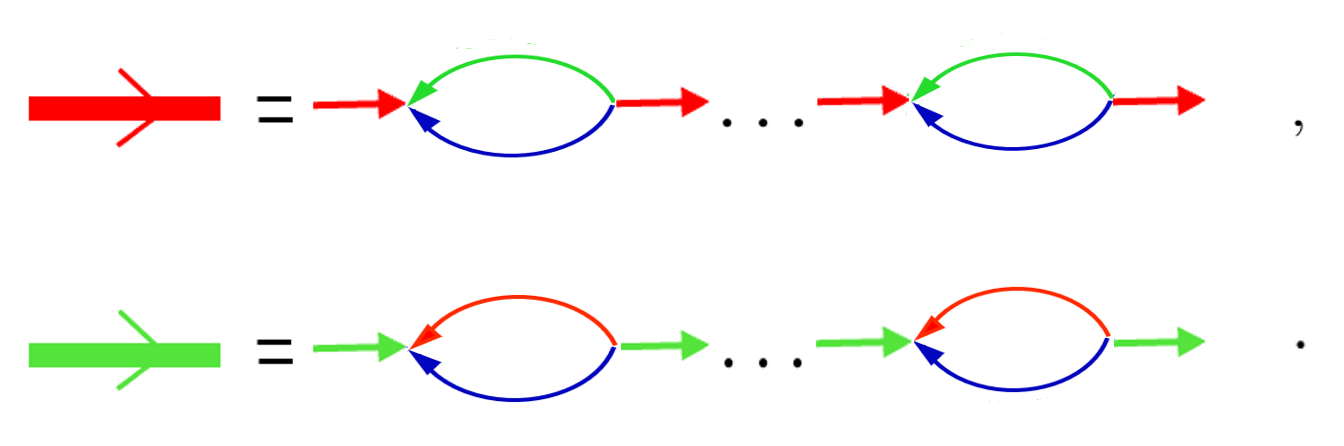}
\end{eqnarray*}

By means of the keystones operators, connected tree and loop operators~$\mathcal{K}_{\sigma}^{(n)}$ in the ring,
we introduce the RG-completed Aristotelian tensor model from (\ref{rainbowrg})
\begin{eqnarray}\label{aristotelian}
Z_{A}&=&\int dA d\bar A \exp(-\mu \Tr A\bar A+t_1^{(1)}\mathcal{K}_1^{(1)}+\sum_{{\textcolor{red}{k}}=2}^\infty {\textcolor{red}{t^{(k)}_{k}}}\mathcal{K}_{{\textcolor{red}{k}}}^{\textcolor{red}{(k)}}
+\sum_{{\textcolor{green}{k}}=2}^\infty \textcolor{green}{t^{(k)}_{k}}\mathcal{K}^{\textcolor{green}{(k)}}_{\textcolor{green}{k}}
+\sum_{{\textcolor{blue}{k}}=2}^\infty \textcolor{blue}{t^{(k)}_{k}}\mathcal{K}^{\textcolor{blue}{(k)}}_{\textcolor{blue}{k}}\nonumber\\ &&+\sum_{{\textcolor{red}{k_1}},{\textcolor{green}{k_2}}=2}^\infty t_{{\textcolor{red}{k_1}},{\textcolor{green}{k_2}}}^{(\textcolor{red}{k_1}+\textcolor{green}{k_2}-1)}\mathcal{K}_{{\textcolor{red}{k_1}},
{\textcolor{green}{k_2}}}^{(\textcolor{red}{k_1}+\textcolor{green}{k_2}-1)}
+\sum_{{\textcolor{red}{k_1}},{\textcolor{blue}{k_2}}=2}^\infty t_{{\textcolor{red}{k_1}},{\textcolor{blue}{k_2}}}^{(\textcolor{red}{k_1}+\textcolor{blue}{k_2}-1)}\mathcal{K}_{{\textcolor{red}
{k_1}},{\textcolor{blue}{k_2}}}^{(\textcolor{red}{k_1}+\textcolor{blue}{k_2}-1)}\nonumber\\
&&+\sum_{{\textcolor{green}{k_1}},{\textcolor{blue}{k_2}}=2}^\infty t_{{\textcolor{green}{k_1}},{\textcolor{blue}{k_2}}}^{(\textcolor{green}{k_1}+\textcolor{blue}{k_2}-1)}\mathcal{K}_{{\textcolor{green}{k_1}},
{\textcolor{blue}{k_2}}}^{(\textcolor{green}{k_1}+\textcolor{blue}{k_2}-1)}
+\cdots)\nonumber\\
&=&\exp(\frac{1}{\mu }\hat{W}_{3})\cdot 1,
\end{eqnarray}
where
\begin{eqnarray}\label{operator3}
\hat{W}_{3}&=&\dsum_{a=1}^{\infty}\dsum_{\deg\alpha=a}\sum_{k=1}^3
\sum_{\substack{\beta_1,\cdots,\beta_k\\b_1+\cdots+b_k+1=a}}(1-\delta_{a,1})
\Delta_{\alpha}^{\beta_1,\cdots,\beta_k}t_{\alpha}^{(a)}
\frac{\partial}{\partial t_{\beta_1}^{(b_1)}}\cdots\frac{\partial}{\partial t_{\beta_k}^{(b_k)}}
+t_{id\otimes id \otimes id}^{(1)}\textcolor{red}{N_{1}}\textcolor{green}{N_{2}} \textcolor{blue}{N_{3}}\nonumber\\
&&
+\dsum_{a,n=1}^{\infty}\dsum_{\deg\alpha=a}\sum_{\deg\sigma=n}\sum_{\deg\beta=n+a-1}\gamma_{\sigma,\alpha}^{\beta}
 t_{\alpha}^{(a)} t_{\sigma}^{(n)}\frac{\partial}{\partial t_{\beta}^{(n+a-1)}},
\end{eqnarray}
$\alpha$, $\sigma$ and $\beta$ are taken from the indices of connected operators in the ring.

For the Aristotelian tensor model  (\ref{aristotelian}),
the Virasoro constraint operators in (\ref{Vcons}) become
\begin{eqnarray}
L_m=(-\frac{1}{\mu})\hat{W}_{3}^m(\hat{W}_{3}-\mu\hat{D}_{3}),
\end{eqnarray}
where $\hat{D}_{3}$ is given by (\ref{operatorD}) with $r=3$ in which the index
$\alpha$ is an element of the double coset $\mathcal{S}_n^3= S_n\backslash S_n^{\otimes 3}/S_n$.

Since the coefficients
$P^{\tau(\alpha_1),\cdots,\tau(\alpha_i)}$ in (\ref{corrf}) follow from the precise expression of
$\hat{W}^m$, we may give the exact correlators from (\ref{corrf}).
In particular,
\begin{eqnarray}
\langle (\mathcal{K}_{1} )^{i} \rangle=\frac{1}{\mu}(\mathcal{N}_3+i-1)\langle (\mathcal{K}_{1} )^{i-1} \rangle=\frac{1}{\mu^i}\prod_{j=0}^{i-1}(\mathcal{N}_3+j),
\end{eqnarray}
where $\mathcal{N}_3=\textcolor{red}{N_{1}}\textcolor{green}{N_{2}} \textcolor{blue}{N_{3}}$.

Let us give the correlators $\langle \mathcal{K}_1 \mathcal{K}_1\rangle$ and
$\langle \mathcal{K}_1 \mathcal{K}_1 \mathcal{K}_1\rangle$
and represent these correlators graphically as follows:
\begin{eqnarray}
%&&\langle K_1 K_1\rangle=\frac{\mathcal{N}_3^2}{\mu^2}+\frac{\mathcal{N}_3}{\mu^2}\nonumber\\
%&&\includegraphics[height=2cm]{K1k1-1}\includegraphics[height=2cm]{K1k1-2}
&&\includegraphics[height=1.9cm]{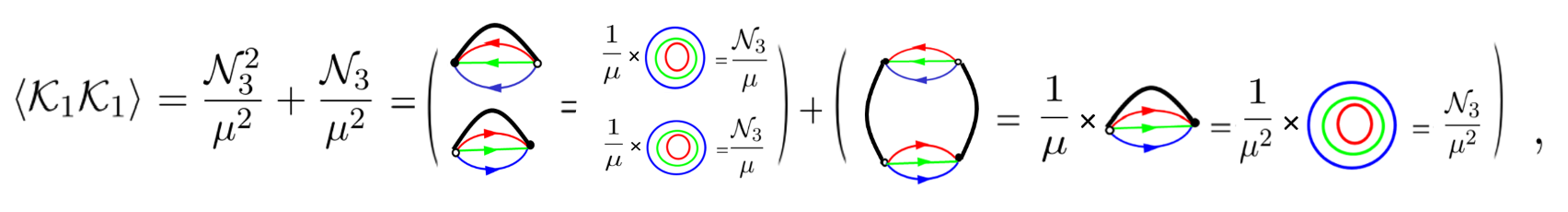}
\nonumber\\
&&\langle \mathcal{K}_1 \mathcal{K}_1 \mathcal{K}_1\rangle
=\frac{\mathcal{N}_3^3}{\mu^3}+3\frac{\mathcal{N}_3^2}{\mu^3}+2\frac{\mathcal{N}_3}{\mu^3}\nonumber\\
&&
\includegraphics[height=2.45cm,width=3.75cm]{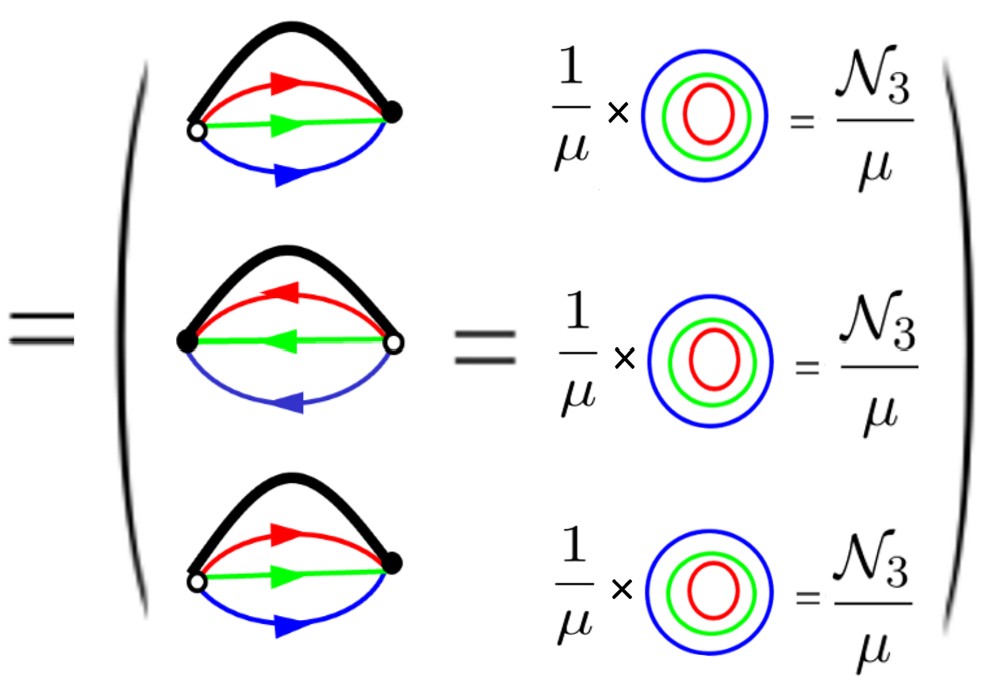}\includegraphics[height=2.25cm]{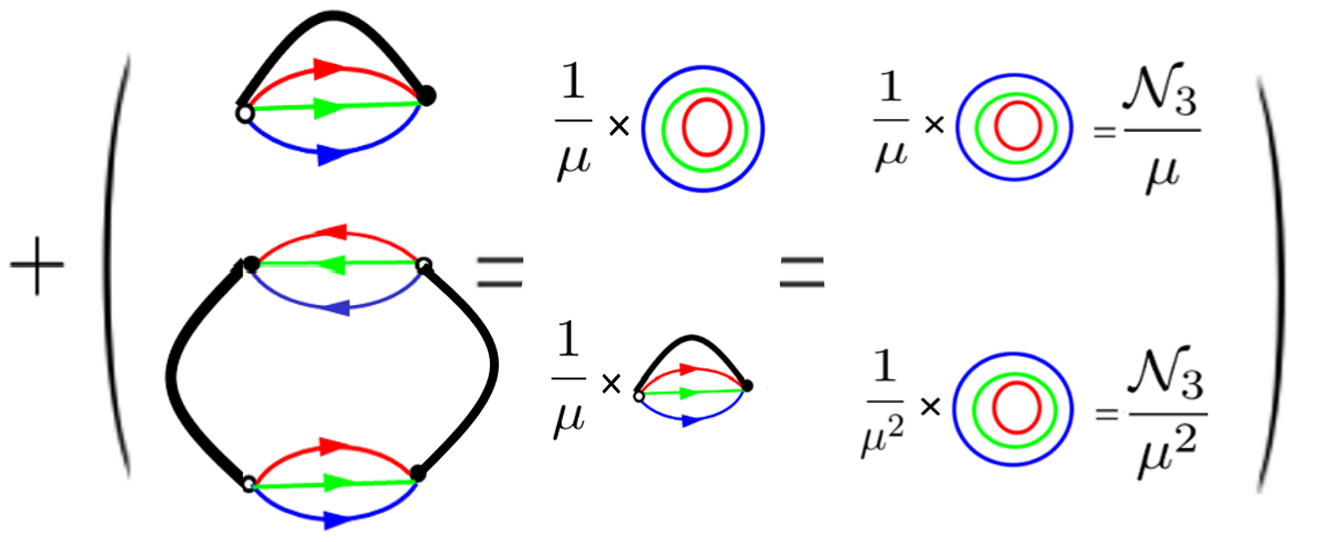}\nonumber\\
&&
\includegraphics[height=2.25cm]{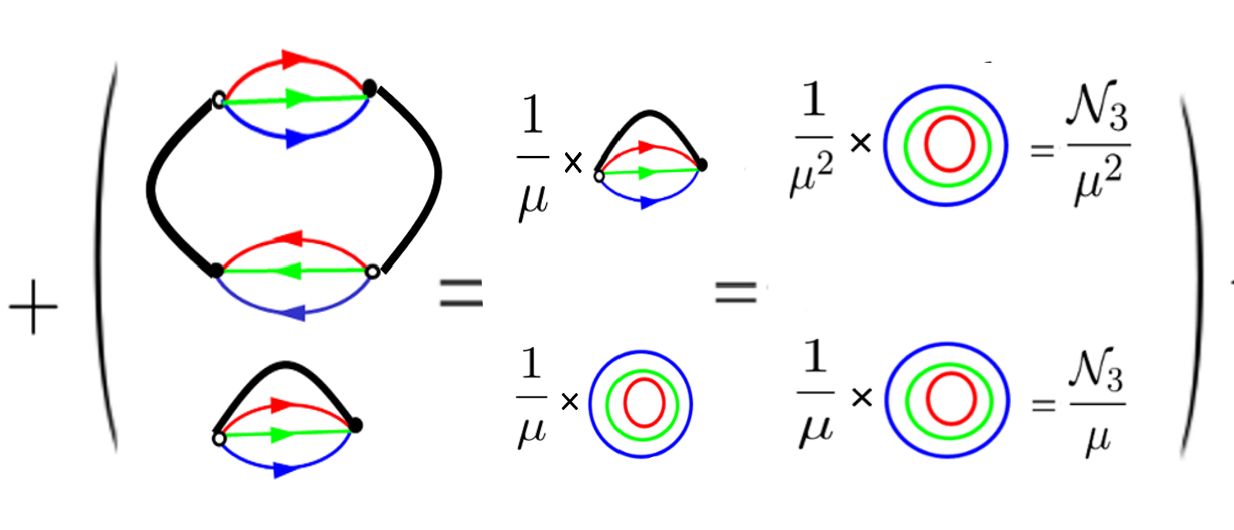}\includegraphics[height=2cm]{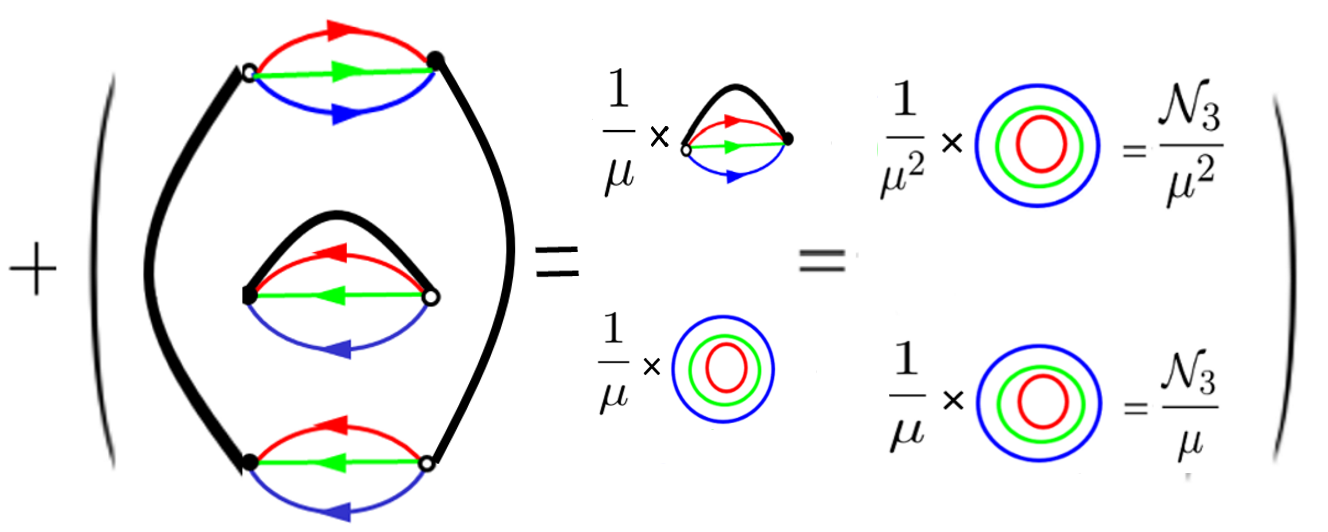}\nonumber\\
&&
\includegraphics[height=1.75cm]{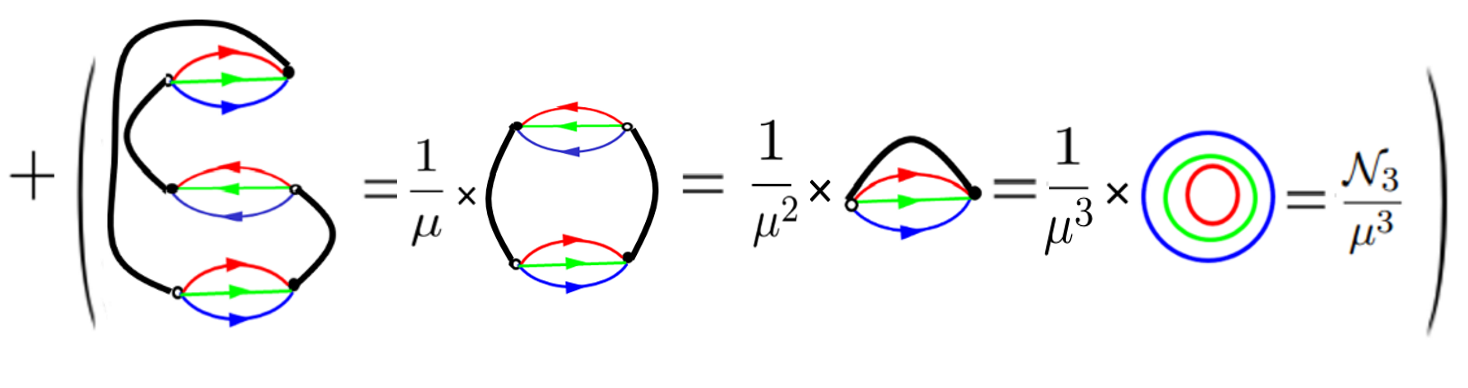}\includegraphics[height=1.85cm]{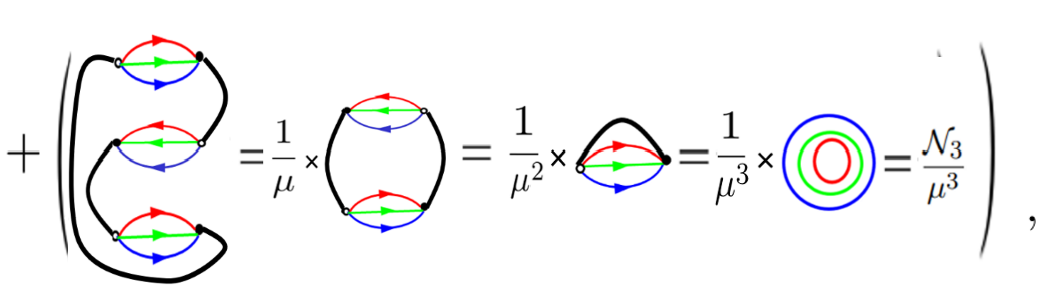}
\end{eqnarray}
where the thick line depicts the Feynman propagator and each propagator gives a factor $\frac{1}{\mu}$,
the red, green and blue circles represent $\textcolor{red}{N_1}$, $\textcolor{green}{N_2}$ and
$\textcolor{blue}{N_3}$, respectively.

By calculating $\hat{W}_3^i$, $i=1,...,4$, we obtain the correlators which
have been derived in~\cite{ItoyamaNPB2018}. In the following, we give two correlators by calculating  $\hat{W}_3^5$
\begin{eqnarray}
 && \left\langle \mathcal{K}_{id\otimes(12345)\otimes id}\right\rangle =\frac{1}{\mu^{5}}[\mathcal{N}_3(15{\textcolor{red}{N_1}}^2{\textcolor{blue}{N_3}}^2
 +15{\textcolor{green}{N_2}}^{2}+{\textcolor{red}{N_1}}^4{\textcolor{blue}{N_3}}^4+{\textcolor{green}{N_2}}^{4}+8)\nonumber\\
&&\qquad\qquad\qquad\qquad\quad
+10\mathcal{N}_3^2({\textcolor{red}{N_1}}^2{\textcolor{blue}{N_3}}^2
 +{\textcolor{green}{N_2}}^{2}+4)+20\mathcal{N}_3^3],
 \nonumber\\
&& \left\langle \mathcal{K}_{id\otimes(12345)\otimes(12)}\right\rangle
 =\frac{1}{\mu^{5}}[\mathcal{N}_3^2({\textcolor{red}{N_1}}^2{\textcolor{blue}{N_3}}^2
 +9{\textcolor{red}{N_1}}^{2}{\textcolor{blue}{N_3}}+14{\textcolor{red}{N_1}}{\textcolor{green}{N_2}}
 +6{\textcolor{green}{N_2}}^2{\textcolor{blue}{N_3}}+20{\textcolor{blue}{N_3}})\nonumber\\
 && \qquad\qquad\qquad\qquad\quad +6\mathcal{N}_3^3{\textcolor{blue}{N_3}}+\mathcal{N}_3({\textcolor{red}{N_1}}^4{\textcolor{blue}{N_3}}^3
+3{\textcolor{red}{N_1}}^{2}{\textcolor{blue}{N_3}}^3+12{\textcolor{red}{N_1}}^2{\textcolor{blue}{N_3}}+20{\textcolor{red}{N_1}}{\textcolor{green}{N_2}}\nonumber\\
 && \qquad\qquad\qquad\qquad\quad
 +4{\textcolor{red}{N_1}}{\textcolor{green}{N_2}}^3
 +15{\textcolor{green}{N_2}}^2{\textcolor{blue}{N_3}}+{\textcolor{green}{N_2}}^4{\textcolor{blue}{N_3}}
 +8{\textcolor{blue}{N_3}})].\nonumber\\
 \end{eqnarray}

\section{Correlators in the $r=4$ rainbow tensor model}
In the rainbow tensor model with the tensor~$A_{\textcolor{red}{i}}^{{\textcolor{green}{j_1}},{\textcolor{blue}{j_2}},{\textcolor{yellow}{j_3}}}$ of rank~$r=4$,
the keystone operators are
\begin{eqnarray}
&&\mathcal{K}_{\textcolor{red}{2}}=\mathcal{K}_{ (12) \otimes id \otimes id\otimes id}
=A_{\textcolor{red}{i}^{(1)}}^{{\textcolor{green}{j_1}^{(1)}},{\textcolor{blue}{j_2}^{(1)}},{\textcolor{yellow}{j_3}^{(1)}}}
\bar{A}_{{\textcolor{green}{j_1}^{(1)}},{\textcolor{blue}{j_2}^{(1)}},{\textcolor{yellow}{j_3}^{(1)}}}^{\textcolor{red}{i}^{(2)}}
A_{\textcolor{red}{i}^{(2)}}^{{\textcolor{green}{j_1}^{(2)}},{\textcolor{blue}{j_2}^{(2)}},{\textcolor{yellow}{j_3}^{(2)}}}
\bar{A}_{{\textcolor{green}{j_1}^{(2)}},{\textcolor{blue}{j_2}^{(2)}},{\textcolor{yellow}{j_3}^{(2)}}}^{\textcolor{red}{i}^{(1)}}
\nonumber\\
&&\includegraphics[height=2cm]{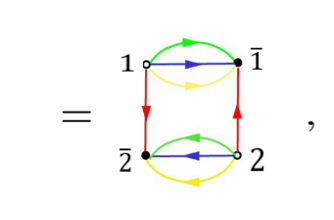}
\end{eqnarray}
\begin{eqnarray}
&&\mathcal{K}_{\textcolor{green}{2}}=\mathcal{K}_{ id \otimes(12)\otimes id\otimes id}=A_{\textcolor{red}{i}^{(1)}}^{{\textcolor{green}{j_1}^{(1)}},{\textcolor{blue}{j_2}^{(1)}},{\textcolor{yellow}{j_3}^{(1)}}}
\bar{A}_{{\textcolor{green}{j_1}^{(2)}},{\textcolor{blue}{j_2}^{(1)}},{\textcolor{yellow}{j_3}^{(1)}}}^{\textcolor{red}{i}^{(1)}}
A_{\textcolor{red}{i}^{(2)}}^{{\textcolor{green}{j_1}^{(2)}},{\textcolor{blue}{j_2}^{(2)}},{\textcolor{yellow}{j_3}^{(2)}}}
\bar{A}_{{\textcolor{green}{j_1}^{(1)}},{\textcolor{blue}{j_2}^{(2)}},{\textcolor{yellow}{j_3}^{(2)}}}^{\textcolor{red}{i}^{(2)}}
\nonumber\\&&\includegraphics[height=2cm]{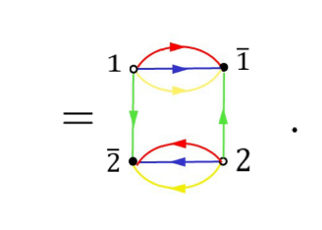}
\end{eqnarray}
The ring generated by the keystone operators  contains
the tree and loop operators, red circles~$\mathcal{K}_{\textcolor{red}{n}}=\mathcal{K}_{ (12\cdots n)  \otimes id\otimes id\otimes id}$, green circles
$\mathcal{K}_{\textcolor{green}{n}}=\mathcal{K}_{id \otimes (12\cdots n)\otimes id\otimes id}$
and disconnected collections of the red and green circles, respectively.

In similarity with the case of Aristotelian tensor model, the tree and loop operators can also be constructed.
In figures.3 and 4, we draw some tree and loop operators involving chains with both
$\mathcal{K}_{\textcolor{red}{2}}$ and $\mathcal{K}_{\textcolor{green}{2}}$, respectively.
\begin{figure}[H]
\centering
\begin{minipage}[c]{0.5\textwidth}
\centering
\includegraphics[height=2cm]{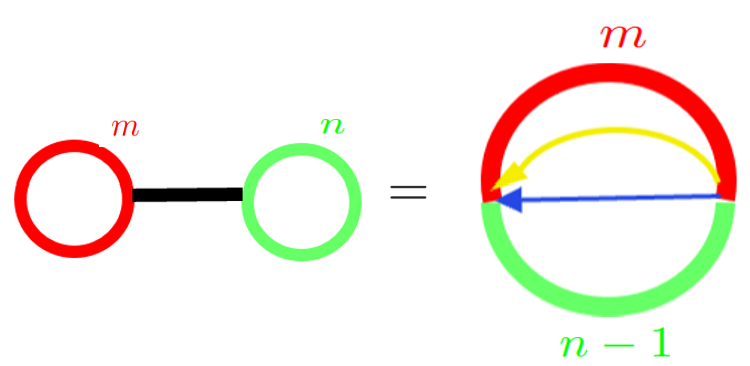}
%${\textcolor{red}{\mathcal{K}_n}}$}
%\label{tu7}
\end{minipage}%
\begin{minipage}[c]{0.5\textwidth}
\centering
\includegraphics[height=2cm]{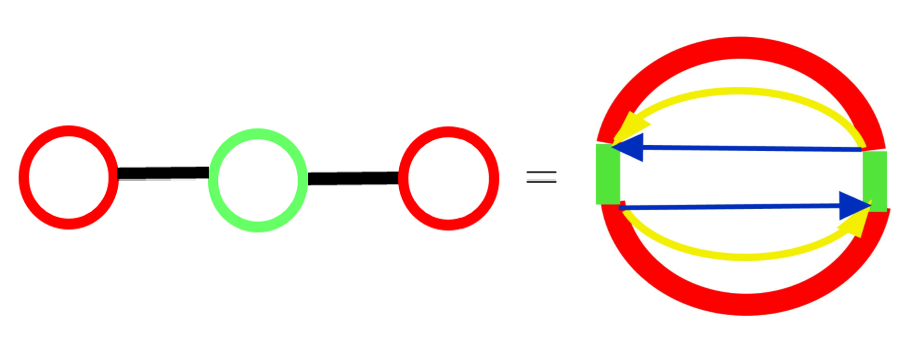}
\end{minipage}
\caption{Diagrams of tree operators.}
\label{treer=4}
\end{figure}

\begin{figure}[H]
\centering
\begin{minipage}[c]{0.5\textwidth}
\centering
\includegraphics[height=1.7cm]{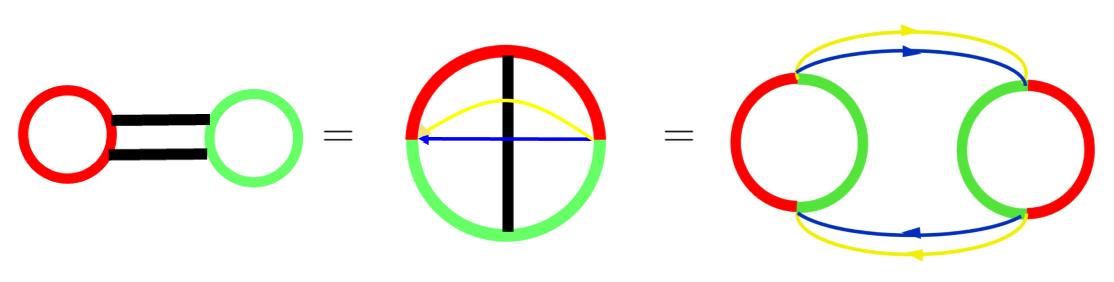}
%${\textcolor{red}{\mathcal{K}_n}}$}
%\label{tu7}
\end{minipage}%
\begin{minipage}[c]{0.5\textwidth}
\centering
\includegraphics[height=2cm]{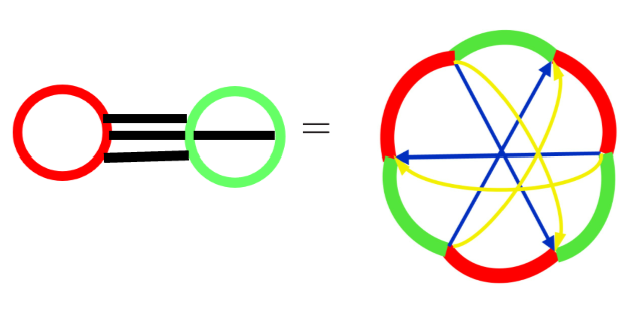}
\end{minipage}
\caption{Diagrams of loop operators.}
\label{loopr=4}
\end{figure}
Corresponding to figures.\ref{alltree}(a), (b) and figures.\ref{loopr=3} (a), (b), we can draw the similar
tree and loop operators made from $\mathcal{K}_{\textcolor{red}{2}}$ or $\mathcal{K}_{\textcolor{green}{2}}$
alone.

Note that here the thick red and green circles are
\begin{eqnarray*}
\includegraphics[height=3cm]{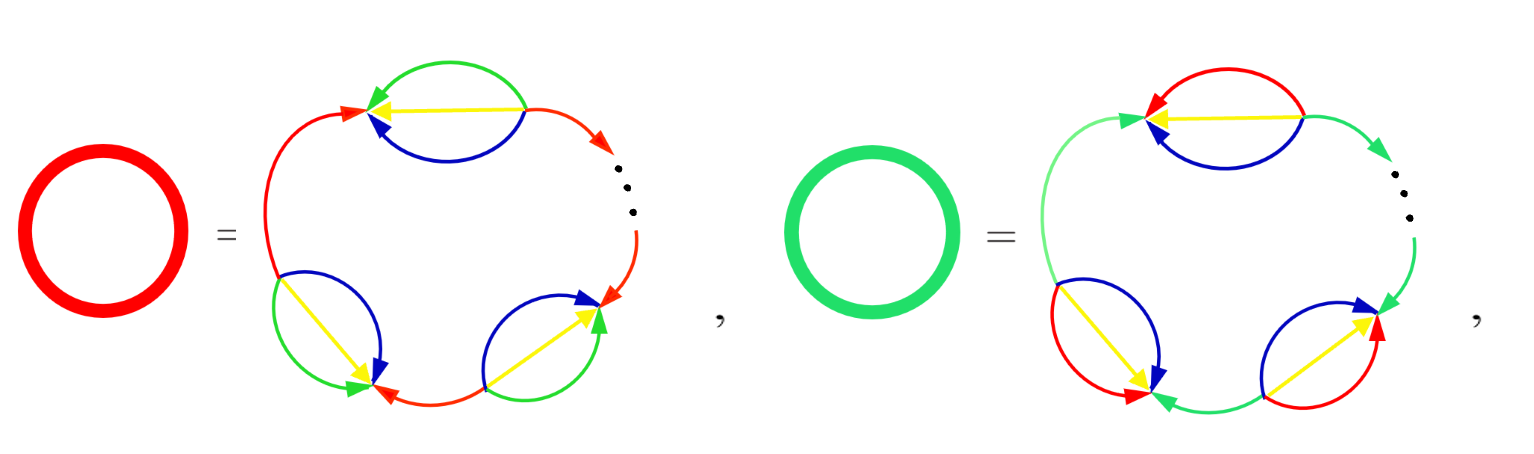}
\end{eqnarray*}
and the thick red and green lines are
\begin{eqnarray*}
\includegraphics[height=3cm]{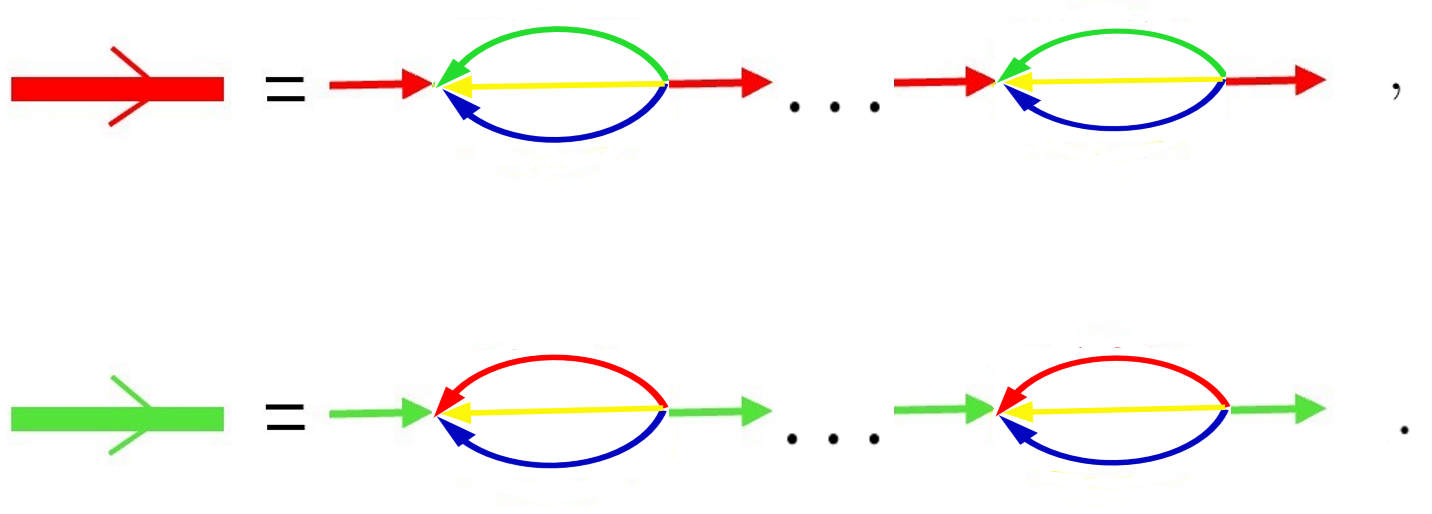}
\end{eqnarray*}

In terms of the connected operators $\mathcal{K}_{\sigma}^{(n)}$, we write the rainbow tensor model (\ref{rainbowrg}) with $r=4$ as
\begin{eqnarray}\label{r4rtm}
Z_{4}&=&\int dA d\bar A \exp(-\mu \Tr A\bar A+t_1^{(1)}\mathcal{K}_1^{(1)}
+\sum_{{\textcolor{red}{k}}=2}^\infty \textcolor{red}{t^{(k)}_{k}}\mathcal{K}_{\textcolor{red}{k}}^{\textcolor{red}{(k)}}
+\sum_{{\textcolor{green}{k}}=2}^\infty \textcolor{green}{t^{(k)}_{k}}\mathcal{K}^{\textcolor{green}{(k)}}_{\textcolor{green}{k}}+\cdots)\nonumber\\
&=&\exp(\frac{1}{\mu }\hat{W}_{4})\cdot 1,
\end{eqnarray}
where the operator~$\hat{W}_4$ is given by
\begin{eqnarray}\label{operator4}
\hat{W}_{4}&=&\dsum_{a,n=1}^{\infty}\dsum_{\deg\alpha=a}\sum_{\deg\sigma=n}\sum_{\deg\beta=n+a-1}\gamma_{\sigma,\alpha}^{\beta}
 t_{\alpha}^{(a)} t_{\sigma}^{(n)}\frac{\partial}{\partial t_{\beta}^{(n+a-1)}}+t_{id\otimes id\otimes id\otimes id}^{(1)}\textcolor{red}{N_{1}}
 \textcolor{green}{N_{2}} \textcolor{blue}{N_{3}}\textcolor{yellow}{N_{4}}\nonumber\\
 &&+\dsum_{a=1}^{\infty}\dsum_{\deg\alpha=a}\sum_{k=1}^4\sum_{\substack{\beta_1,\cdots,\beta_k\\b_1
 +\cdots+b_k+1=a}}(1-\delta_{a,1})\Delta_{\alpha}^{\beta_1,\cdots,\beta_k}t_{\alpha}^{(a)}
\frac{\partial}{\partial t_{\beta_1}^{(b_1)}}\cdots\frac{\partial}{\partial t_{\beta_k}^{(b_k)}}.
\end{eqnarray}

For the partition function (\ref{r4rtm}),
the Virasoro constraint operators in (\ref{Vcons}) are
\begin{eqnarray}
L_m=(-\frac{1}{\mu})\hat{W}_{4}^m(\hat{W}_{4}-\mu\hat{D}_{4}),
\end{eqnarray}
where $\hat{D}_{4}$ is given by (\ref{operatorD}) with $r=4$ in which the index
$\alpha$ is an element of the double coset $\mathcal{S}_n^4= S_n\backslash S_n^{\otimes 4}/S_n$.

We may give the exact correlators from (\ref{corrf}), where the coefficients
$P^{\tau(\alpha_1),\cdots,\tau(\alpha_i)}$ in (\ref{corrf}) follow from the power of
$\hat{W}_{4}$ (\ref{operator4}).
The special correlators $\langle (\mathcal{K}_{1} )^{i} \rangle$ are given by
(\ref{spcorrf}) with $\mathcal{N}_4=\textcolor{red}{N_{1}}\textcolor{green}{N_{2}} \textcolor{blue}{N_{3}}\textcolor{yellow}{N_{4}}$.

Let us list some correlators (\ref{corrf}) by calculating $\hat{W}_{4}^i$, $i=1,2,3$, and represent them graphically
with the same rules as the case of Aristotelian tensor model.
Noted that the thick ($4$-colored) lines depict the Feynman propagators, and the yellow circles represent $\textcolor{yellow}{N_4}$.

(\rmnum{1})
\begin{eqnarray}\label{k1oper4}
%\left\langle \mathcal{K}_{1}\right\rangle &&=\left\langle \mathcal{K}^{(1)}_{id\otimes id\otimes id\otimes id}\right\rangle=\frac{\mathcal{N}_4}{\mu}\nonumber\\
%&&\includegraphics[height=2cm]{4-K1-1}
\includegraphics[height=1.15cm]{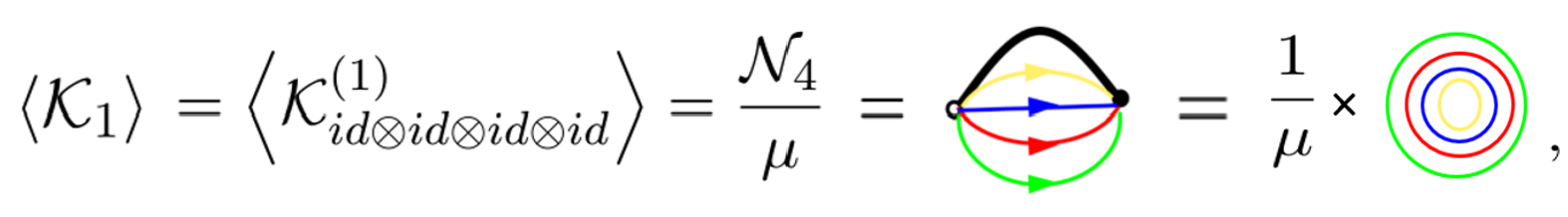}
\end{eqnarray}
there is only one possible attachment
of the Feynman propagators to the operator~$\mathcal{K}_{1}$, giving the result~$\frac{\mathcal{N}_4}{\mu}$.

(\rmnum{2})
\begin{eqnarray*}\label{k2oper4}
  && \left\langle\mathcal{K}_{ \textcolor{red}{2}}\right\rangle=
   \left\langle \mathcal{K}_{id \otimes (12)\otimes (12) \otimes(12)}^{(2)}\right\rangle
   =\frac{\textcolor{red}{N_1}\mathcal{N}_4} {\mu^{2}}+\frac{\textcolor{green}{N_2}\textcolor{blue}{N_3}\textcolor{yellow}{N_4}\mathcal{N}_4}{\mu^2}
   \nonumber\\
&& \includegraphics[height=2cm]{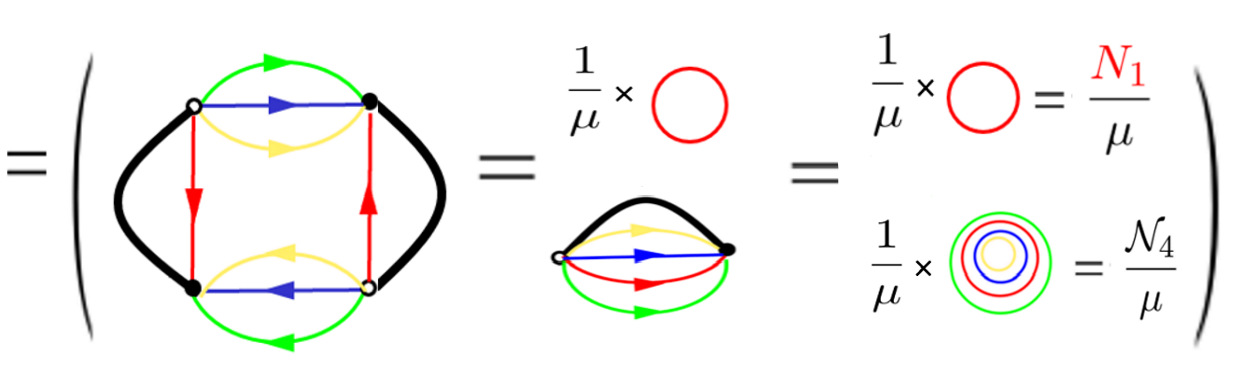}\includegraphics[height=2cm]{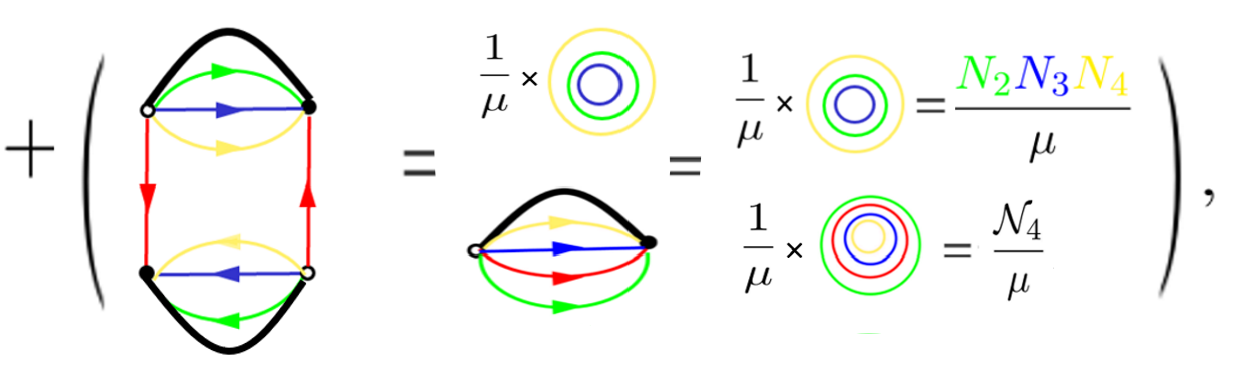}
\end{eqnarray*}
\begin{eqnarray}\label{k2idid1212}
&&\left\langle \mathcal{K}_{id \otimes id\otimes (12) \otimes(12)}^{(2)}\right\rangle=
  \frac{\textcolor{red}{N_1}\textcolor{green}{N_2}\mathcal{N}_4}{\mu^{2}}
  +\frac{\textcolor{blue}{N_3}\textcolor{yellow}{N_4} \mathcal{N}_4}{\mu^{2}}\nonumber\\
  &&\includegraphics[height=2cm]{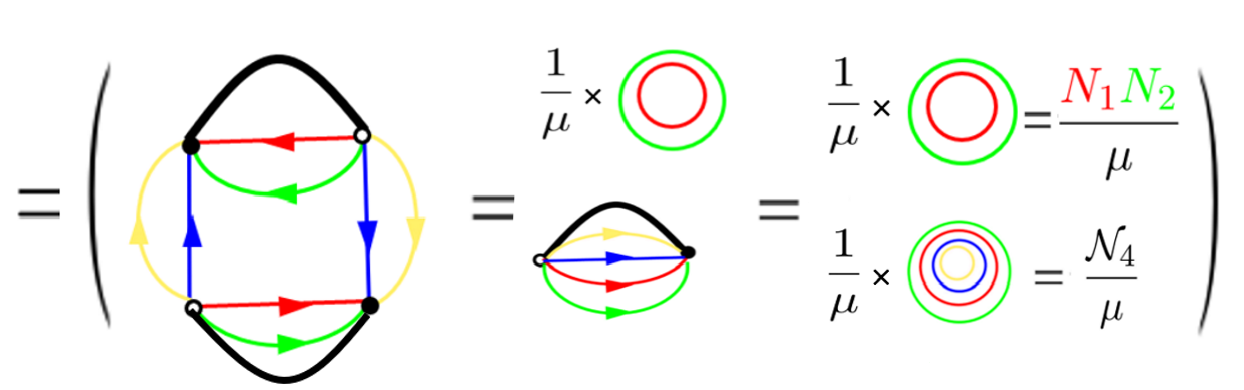}\includegraphics[height=2cm]{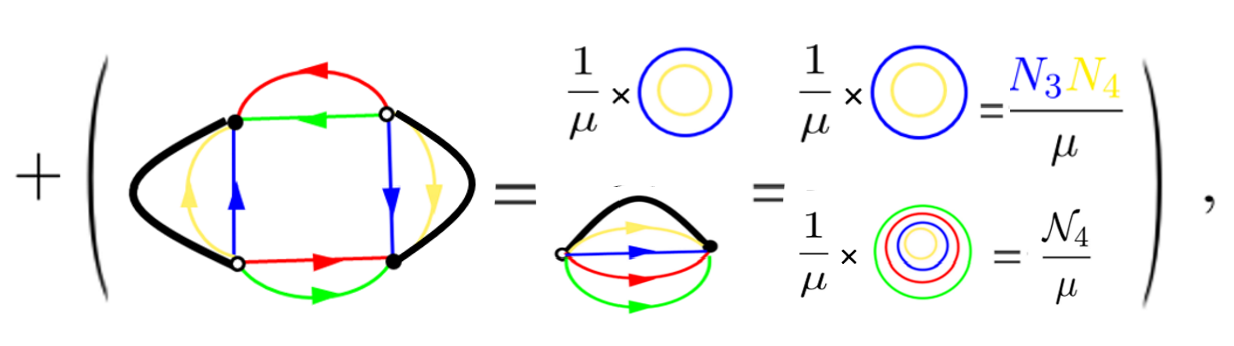}\nonumber\\
%\end{eqnarray}
%\begin{eqnarray}
 \nonumber\\
 && \left\langle \mathcal{K}_{1}\mathcal{K}_{1}\right\rangle=
   \left\langle \mathcal{K}_{id\otimes id\otimes id\otimes id}^{(1)}\mathcal{K}_{id\otimes id\otimes id\otimes id}^{(1)}\right\rangle=
   \frac{ \mathcal{N}_4(\mathcal{N}_4+1)}{\mu^{2}}\nonumber\\
   &&\includegraphics[height=2cm]{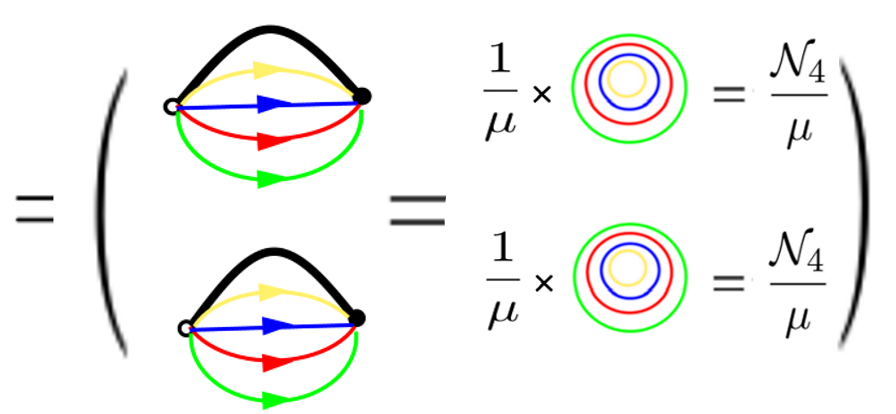}\includegraphics[height=2cm]{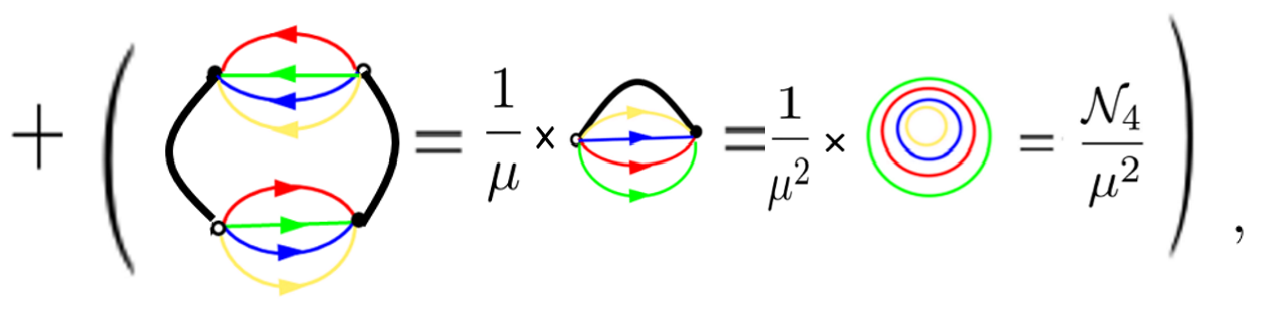}\nonumber\\
%\end{eqnarray}
%\begin{eqnarray}
 \nonumber\\
  &&\left\langle\mathcal{K}_{ \textcolor{green}{2}}\right\rangle=
   \left\langle \mathcal{K}_{id\otimes (12)\otimes id\otimes id}^{(2)}\right\rangle
   =\frac{\textcolor{green}{N_2}\mathcal{N}_4}{\mu^{2}}
   +\frac{\textcolor{red}{N_1}\textcolor{blue}{N_3}\textcolor{yellow}{N_4}\mathcal{N}_4}{\mu^2}
\nonumber\\
&&\includegraphics[height=2cm]{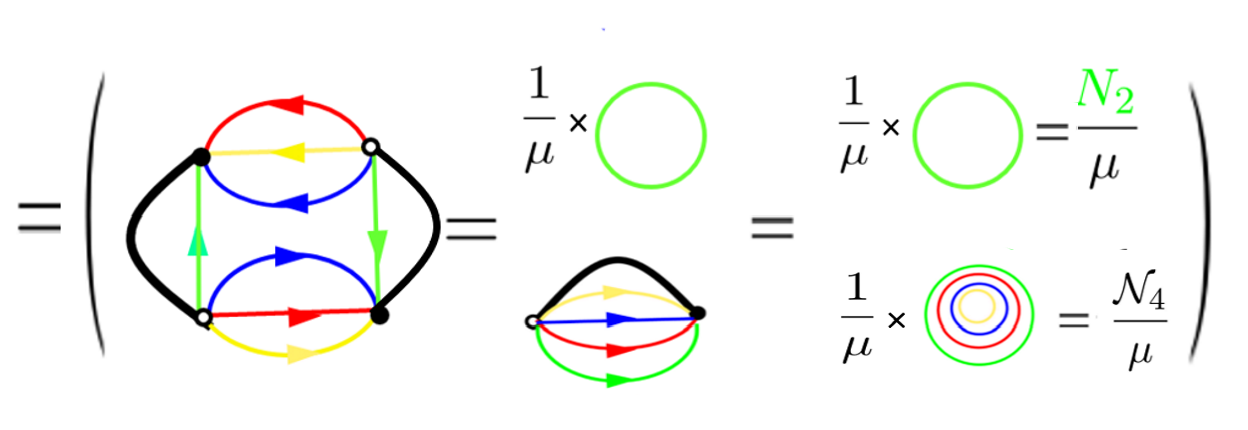}\includegraphics[height=1.8cm]{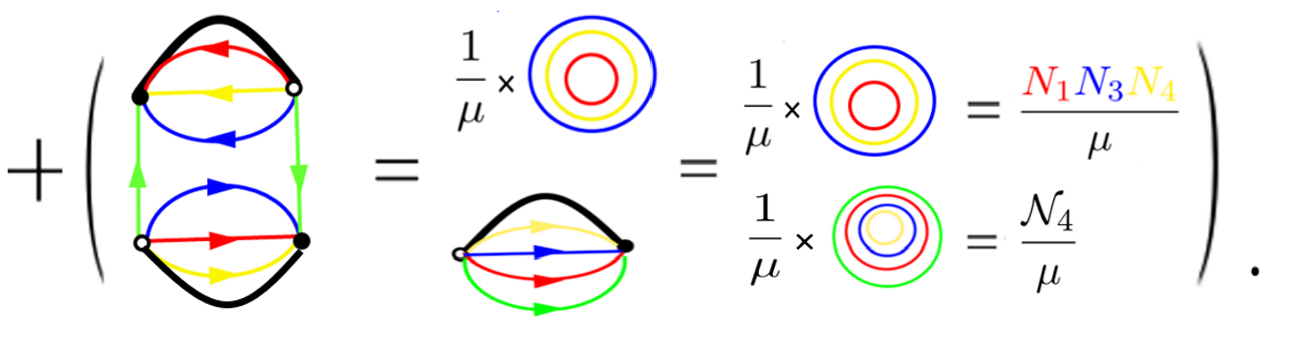}
\end{eqnarray}

(\rmnum{3})
\begin{eqnarray}
  &&\left\langle\mathcal{K}_{\textcolor{red}{3}}\right\rangle=\left\langle \mathcal{K}_{id \otimes (123)\otimes(123)\otimes(123)}^{(3)}\right\rangle\nonumber\\
   &&=
   \frac{1 }{\mu^{3}}(\textcolor{red}{N_1}\cdot \textcolor{red}{N_1}\mathcal{N}_4+2\textcolor{red}{N_1}\cdot \textcolor{green}{N_2} \textcolor{blue}{N_3}\textcolor{yellow}{N_4}\mathcal{N}_4
   +\textcolor{green}{N_2} \textcolor{blue}{N_3}\textcolor{yellow}{N_4}\cdot\textcolor{green}{N_2} \textcolor{blue}{N_3}\textcolor{yellow}{N_4}\mathcal{N}_4
   +\mathcal{N}_4^2+\mathcal{N}_4)
   \nonumber\\
   &&\includegraphics[height=2cm]{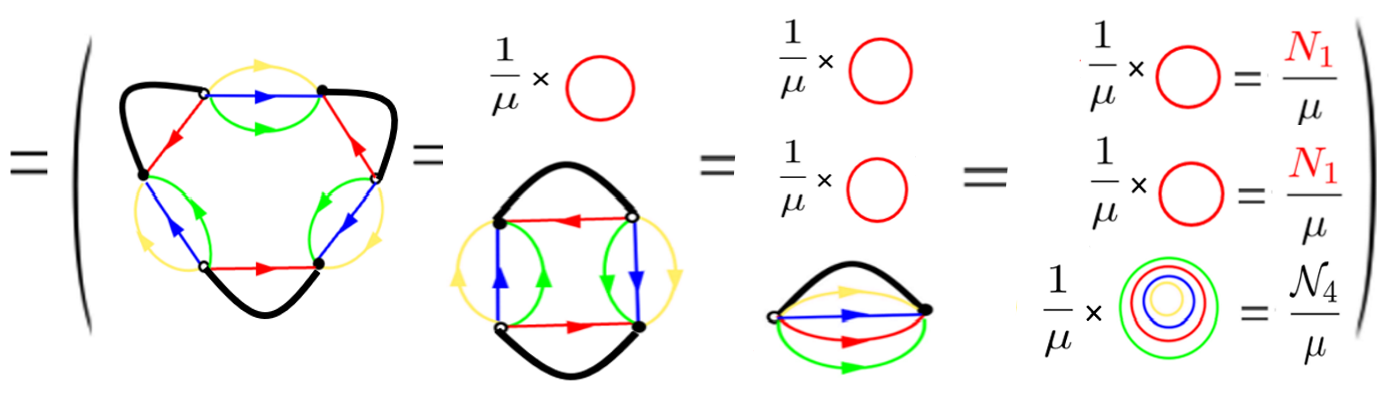}\includegraphics[height=2.25cm]{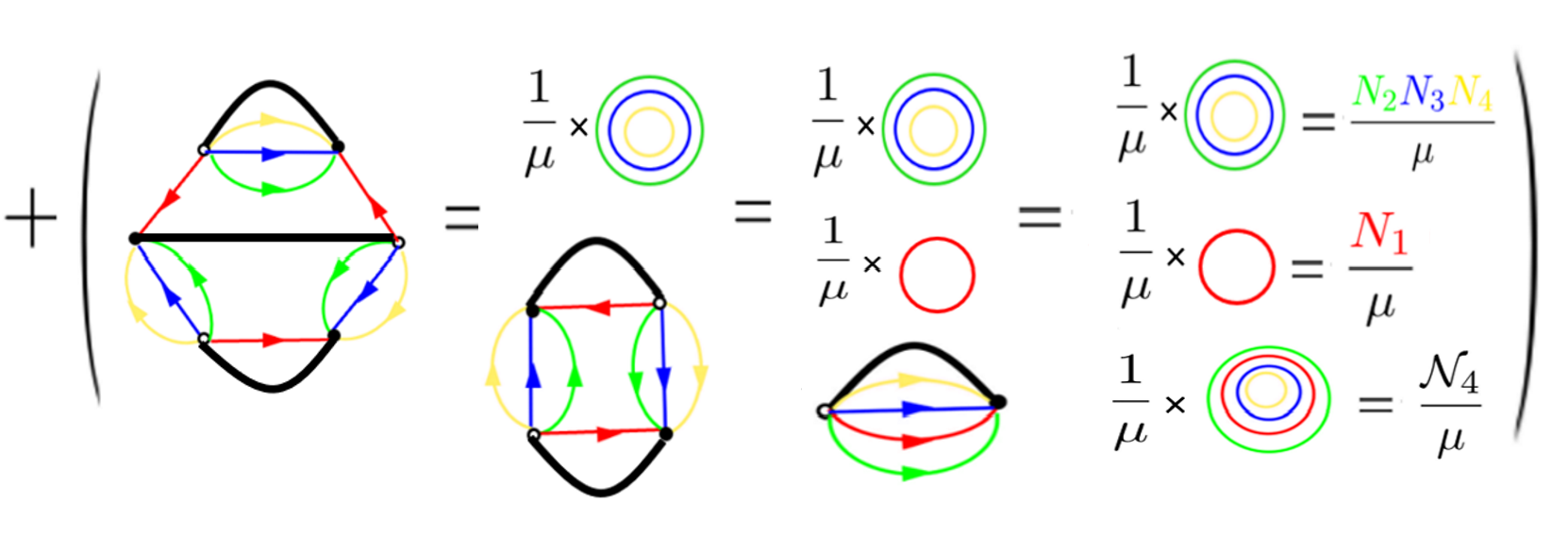}
   \nonumber\\
   &&\includegraphics[height=2.25cm]{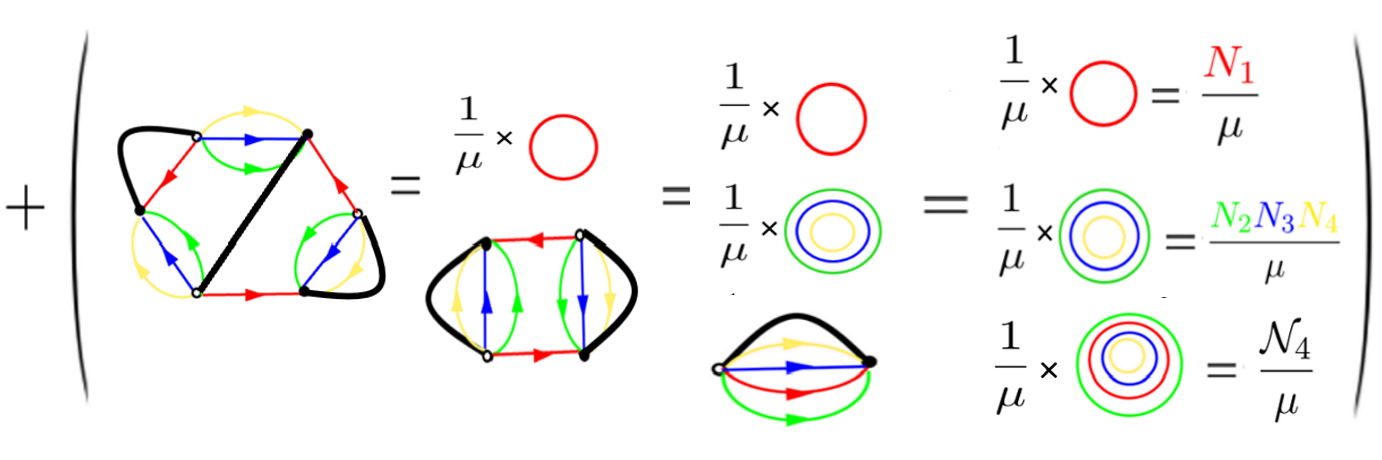}\includegraphics[height=2.25cm]{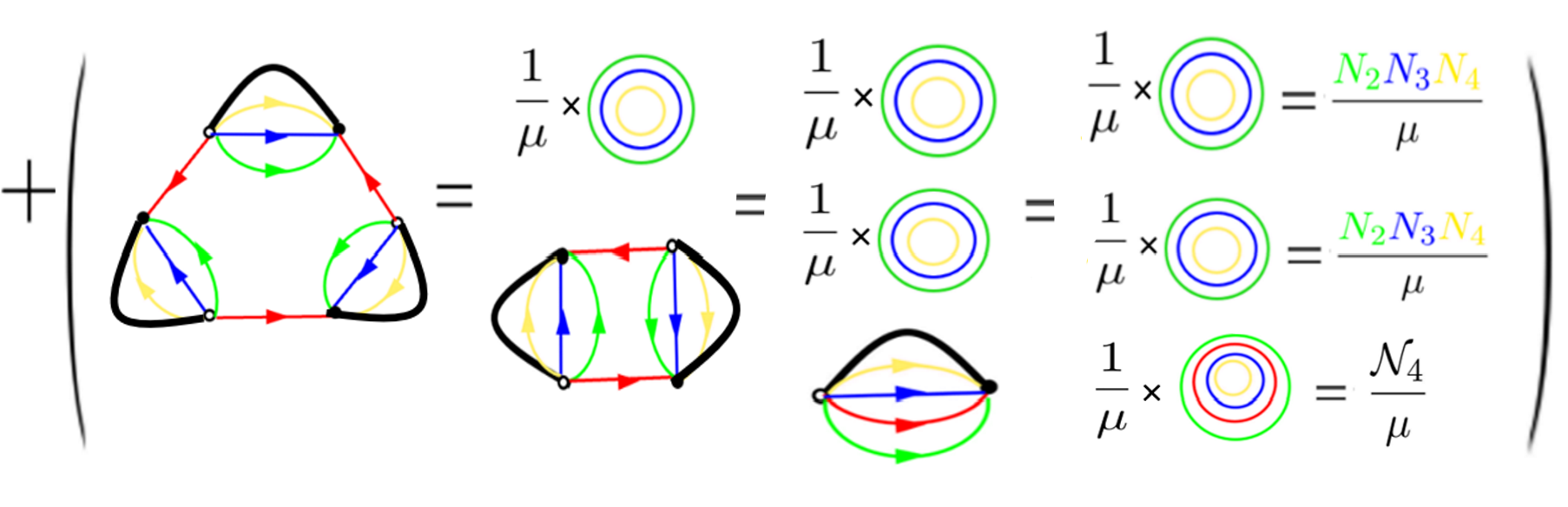}
    \nonumber\\
   &&\includegraphics[height=1.7cm]{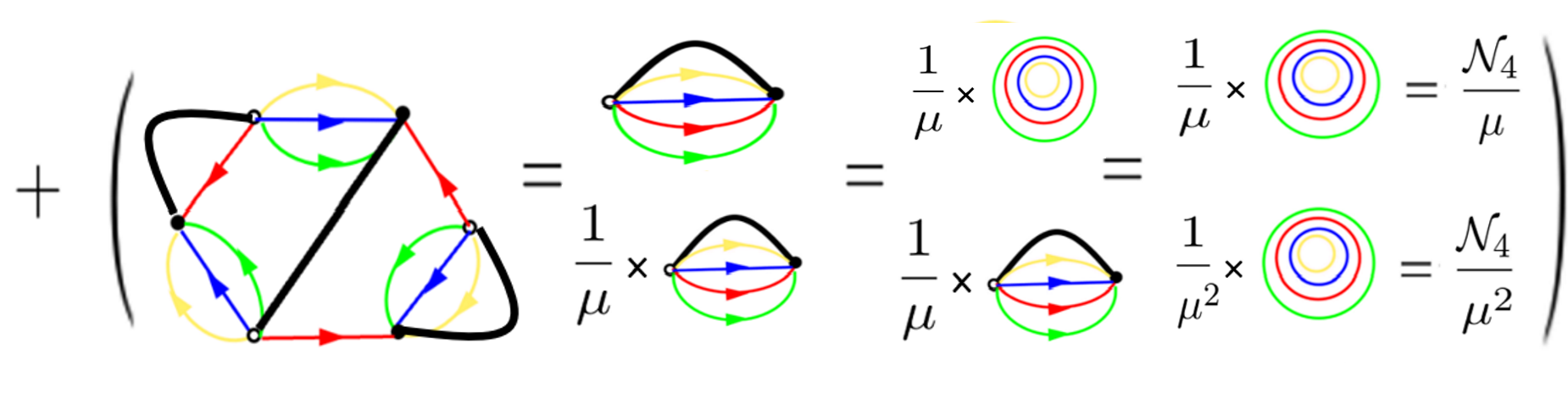}\includegraphics[height=1.7cm]{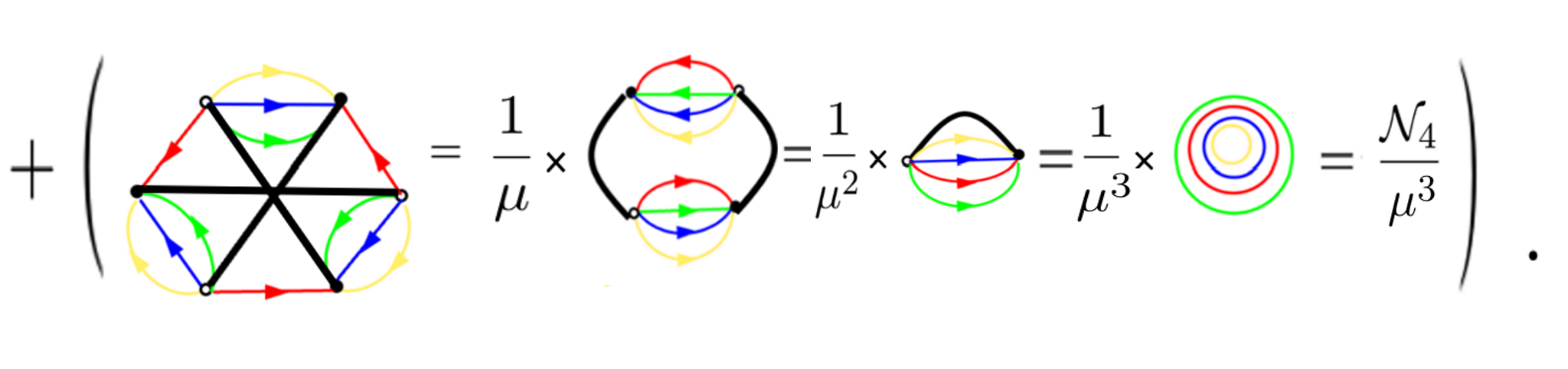}
\end{eqnarray}

\begin{eqnarray}\label{FigF}
\left\langle   \mathcal{K}_{id \otimes (123)\otimes(12)\otimes(12)}^{(3)}\right\rangle
&=&\frac{\mathcal{N}_4}{\mu^3}
(2{\textcolor{red}{N_1}}\cdot{\textcolor{green}{N_2}}+{\textcolor{red}{N_1}}\cdot{\textcolor{red}{N_1}}
{\textcolor{blue}{N_3}}\textcolor{yellow}{N_{4}}
+{\textcolor{green}{N_2}}\cdot{\textcolor{green}{N_2}}{\textcolor{blue}{N_3}}\textcolor{yellow}{N_{4}}
+{\textcolor{blue}{N_3}}\textcolor{yellow}{N_{4}}\nonumber\\
&&+\frac{1}{3}{\textcolor{blue}{N_3}}\textcolor{yellow}{N_{4}}\mathcal{N}_4+\frac{2}{3}{\textcolor{red}
{N_1}}{\textcolor{blue}{N_3}}\textcolor{yellow}{N_{4}}\cdot{\textcolor{green}{N_2}}{\textcolor{blue}{N_3}}
\textcolor{yellow}{N_{4}}
),\nonumber\\
\left\langle   \mathcal{K}_{id \otimes id\otimes(123)\otimes(123)}^{(3)}\right\rangle
&=&\frac{\mathcal{N}_4}{\mu^3}
(\textcolor{red}{N_1}\textcolor{green}{N_2}\cdot\textcolor{red}{N_1}\textcolor{green}{N_2}
+2\textcolor{red}{N_1}\textcolor{green}{N_2}\cdot\textcolor{blue}{N_3}\textcolor{yellow}{N_4}
+\textcolor{blue}{N_3}\textcolor{yellow}{N_4}\cdot\textcolor{blue}{N_3}\textcolor{yellow}{N_4}
+\mathcal{N}_4+1),\nonumber\\
\left\langle   \mathcal{K}_{id \otimes (123)\otimes(23)\otimes(23)}^{(3)}\right\rangle
&=&\frac{\mathcal{N}_4}{\mu^3}
(2{\textcolor{green}{N_2}}\cdot{\textcolor{blue}{N_3}}\textcolor{yellow}{N_{4}}+
{\textcolor{green}{N_2}}\cdot{\textcolor{red}{N_1}}{\textcolor{green}{N_2}}
+{\textcolor{blue}{N_3}}\textcolor{yellow}{N_{4}}\cdot{\textcolor{red}{N_1}}{\textcolor{blue}{N_3}}
\textcolor{yellow}{N_{4}}+{\textcolor{red}{N_1}}\nonumber\\
&&+\frac{2}{3}{\textcolor{red}{N_1}}{\textcolor{green}{N_2}}\cdot{\textcolor{red}{N_1}}{\textcolor{blue}{N_3}}
\textcolor{yellow}{N_{4}}+\frac{1}{3}{\textcolor{red}{N_1}}\mathcal{N}_4),\nonumber\\
\left\langle \mathcal{K}_{id \otimes (123)\otimes(132)\otimes(132)}^{(3)}\right\rangle
&=& \frac{\mathcal{N}_4}{\mu^3}
({\textcolor{red}{N_1}}\cdot\textcolor{red}{N_1}
+{\textcolor{red}{N_1}}\cdot{\textcolor{green}{N_2}}{\textcolor{blue}{N_3}}\textcolor{yellow}{N_{4}}
+{\textcolor{green}{N_2}}\cdot{\textcolor{green}{N_2}}
+{\textcolor{green}{N_2}}\cdot{\textcolor{red}{N_1}}{\textcolor{blue}{N_3}}\textcolor{yellow}{N_{4}}\nonumber\\
&&+{\textcolor{blue}{N_3}}\textcolor{yellow}{N_{4}}\cdot{\textcolor{blue}{N_3}}\textcolor{yellow}{N_{4}}
+{\textcolor{blue}{N_3}}\textcolor{yellow}{N_{4}}\cdot{\textcolor{red}{N_1}}{\textcolor{green}{N_2}}),\nonumber\\
\left\langle \mathcal{K}_{id\otimes (12)\otimes (12)\otimes (12)}^{(2)}\mathcal{K}_{id\otimes id\otimes id\otimes id}^{(1)}\right\rangle
&=&\frac{\mathcal{N}_4}{\mu^3}(2\textcolor{red}{N_1}+2{\textcolor{green}{N_2}}{\textcolor{blue}{N_3}}
\textcolor{yellow}{N_{4}}
+\textcolor{red}{N_1}\mathcal{N}_4+{\textcolor{green}{N_2}}{\textcolor{blue}{N_3}}
\textcolor{yellow}{N_{4}}\mathcal{N}_4),\nonumber\\
\left\langle \mathcal{K}_{id\otimes id\otimes (12)\otimes (12)}^{(2)}\mathcal{K}_{id\otimes id\otimes id\otimes id}^{(1)}\right\rangle
&=&\frac{ \mathcal{N}_4}{\mu^{3}}
(2\textcolor{red}{N_1}\textcolor{green}{N_2}+2\textcolor{blue}{N_3}\textcolor{yellow}{N_4}
+\textcolor{red}{N_1}\textcolor{green}{N_2}\mathcal{N}_4+\textcolor{blue}{N_3}
\textcolor{yellow}{N_4}\mathcal{N}_4)
.
\end{eqnarray}
Due to too many Feynman diagrams for the correlators (\ref{FigF}), we do not present them here.

\section{(Non-Gaussian) Red tensor model}

\subsection{Correlators in the red tensor model}
Let's take~$\sigma$ in (\ref{rainbowrg}) to be the simplest element of the double coset
$\mathcal{S}_n^r=S_n\backslash S_n^{\otimes r}/S_n$, i.e. $(12\cdots n) \otimes id\otimes  \cdots \otimes id$, then
\begin{eqnarray}
\textcolor{red}{\mathcal{K}_{n}}\equiv
\mathcal{K}_\sigma^{(n)}
&=&
\mathcal{K}_{(12\cdots n)\otimes id\otimes\cdots\otimes id}^{(n)}\nonumber\\
&=&A_{\textcolor{red}{i}^{(1)}}^{{\textcolor{green}{j_1}^{(1)}},{\textcolor{blue}{j_2}^{(1)}},\cdots,{j_{r-1}}^{(1)}}
\bar{A}_{{\textcolor{green}{j_1}^{(1)}},{\textcolor{blue}{j_2}^{(1)}},\cdots,{j_{r-1}}^{(1)}}^{\textcolor{red}{i}^{(2)}}
A_{\textcolor{red}{i}^{(2)}}^{{\textcolor{green}{j_1}^{(2)}},{\textcolor{blue}{j_2}^{(2)}},\cdots,{j_{r-1}}^{(2)}}
\bar{A}_{{\textcolor{green}{j_1}^{(2)}},{\textcolor{blue}{j_2}^{(2)}},\cdots,{j_{r-1}}^{(2)}}^{\textcolor{red}{i}^{(3)}}
\nonumber\\
&&
\cdots
A_{\textcolor{red}{i}^{(n-1)}}^{{\textcolor{green}{j_1}^{(n-1)}},{\textcolor{blue}{j_2}^{(n-1)}},\cdots,{j_{r-1}}^{(n-1)}}
\bar{A}_{{\textcolor{green}{j_1}^{(n-1)}},{\textcolor{blue}{j_2}^{(n-1)}},\cdots,{j_{r-1}}^{(n-1)}}^{\textcolor{red}{i}^{(n)}}
\nonumber\\
&&
\cdot A_{\textcolor{red}{i}^{(n)}}^{{\textcolor{green}{j_1}^{(n)}},{\textcolor{blue}{j_2}^{(n)}},\cdots,{j_{r-1}}^{(n)}}
\bar{A}_{{\textcolor{green}{j_1}^{(1)}},{\textcolor{blue}{j_2}^{(1)}},\cdots,{j_{r-1}}^{(1)}}^{\textcolor{red}{i}^{(1)}}.
\end{eqnarray}

Let us take the keystone operator
\begin{eqnarray}\label{keystonetrivial}
&&\includegraphics[height=4.5cm]{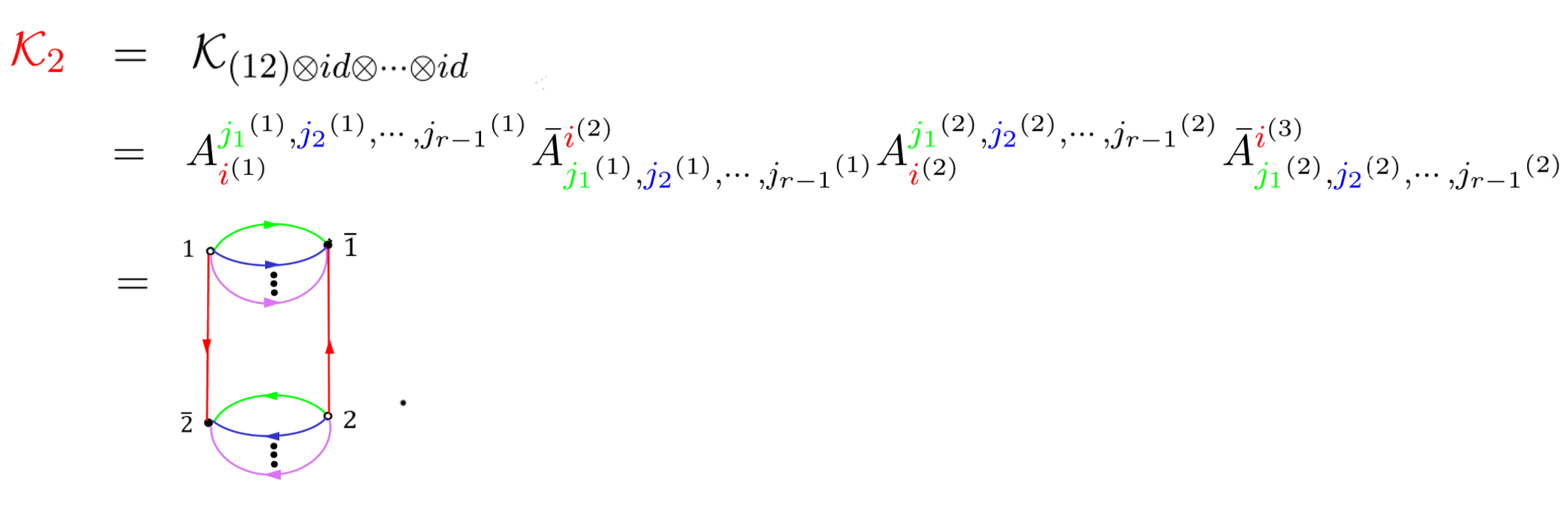}
\end{eqnarray}
Then the tree and loop operators are $\textcolor{red}{{\mathcal{K}_n}}$ and $\prod_{\textcolor{red}{n}}\textcolor{red}{{\mathcal{K}_n}}$, respectively.
Note that the tree operators are connected.

We introduce the red tensor model
\begin{eqnarray}\label{red}
\textcolor{red}{Z}=\int dA d\bar A \exp(-\mu \Tr A\bar A+\sum_{{\textcolor{red}{n}}=1}^\infty {\textcolor{red}{t_n}}{\textcolor{red}{\mathcal{K}_n}}
)
=\exp(\frac{1}{\mu }\textcolor{red}{\hat{W}})\cdot 1,
\end{eqnarray}
where we denote the index~$i$ in~$A_i^{j_1,\ldots,j_{r-1}}$ with color red,
the operator $\textcolor{red}{\hat{W}}$ is given by
\begin{eqnarray}\label{w1}
\textcolor{red}{\hat{W}}&=&\sum_{{\textcolor{red}{b_1,b_2}}=1}^{\infty}({\textcolor{red}{b_1}}+{\textcolor{red}{b_2}}+1){\textcolor{red}{t_{b_1+b_2+1}}}
\frac{\partial}{\partial {\textcolor{red}{t_{b_1}}}}\frac{\partial}{\partial {\textcolor{red}{t_{b_2}}}}
+\sum_{{\textcolor{red}{b_1,b_2}}=1}^{\infty}{\textcolor{red}{b_1b_2t_{b_1}t_{b_2}}}\frac{\partial}{\partial
\textcolor{red}{t_{b_1+b_2-1}}}\nonumber\\
&&+\textcolor{red}{\tilde{\mathcal{N}}_r} \sum_{{\textcolor{red}{b}}=1}^{\infty}({\textcolor{red}{b}}+1){\textcolor{red}{t_{b+1}}}\dfrac{\partial }{\partial {\textcolor{red}{t_b}}}
+\mathcal{N}_r {\textcolor{red}{t_1}},
\end{eqnarray}
$\textcolor{red}{\tilde{\mathcal{N}}_r}=\textcolor{red}{N_{1}}+\textcolor{green}{N_{2}}\textcolor{blue}{N_{3}}\cdots N_{r}$
and $\mathcal{N}_r=\textcolor{red}{N_{1}}\textcolor{green}{N_{2}}\textcolor{blue}{N_{3}}\cdots N_{r}$.

When particularized to the rank 3 tensor case in the partition function (\ref{red}),
the operator (\ref{w1}) reduces to the result derived in Ref.\cite{ItoyamaJHEP2017}.

For the case of the red tensor model (\ref{red}), the Virasoro constraint operators in (\ref{Vcons}) are
\begin{eqnarray}\label{virasorom}
L_m=(-\frac{1}{\mu})\textcolor{red}{\hat{W}}^m(\textcolor{red}{\hat{W}}-\mu\textcolor{red}{\hat{D}}),
\end{eqnarray}
and the correlators (\ref{corrf}) become
\begin{eqnarray}\label{redc}
\langle {\textcolor{red}{\mathcal{K}_{a_1}}} \cdots {\textcolor{red}{\mathcal{K}_{a_i}}} \rangle=\frac{i!}{\mu^{m}m! \lambda_{({\textcolor{red}{a_1}},\cdots, {\textcolor{red}{a_i}})}}
\sum_{\tau}P_{r}^{\tau({\textcolor{red}{a_1}}),\cdots, \tau({\textcolor{red}{a_i}})},
\end{eqnarray}
where $m={\textcolor{red}{a_1}}+\cdots+{\textcolor{red}{a_{i}}}$
and $\textcolor{red}{\hat{D}}=\dsum_{\textcolor{red}{a}=1}^{\infty}\textcolor{red}{at_a}\frac{\partial}{\partial \textcolor{red}{t_a}} $.
Furthermore the correlators $\langle ({\textcolor{red}{\mathcal{K}_{1}}} )^{i} \rangle$ are given by
(\ref{spcorrf}) with $\mathcal{N}_r=\textcolor{red}{N_{1}}\textcolor{green}{N_{2}}\textcolor{blue}{N_{3}}\cdots N_{r}$.

By calculating $\textcolor{red}{\hat{W}}^{i}, i = 1,...,4$ to give
$P_{r}^{\tau({\textcolor{red}{a_1}}),\cdots, \tau({\textcolor{red}{a_i}})}$ in
(\ref{redc}), we obtain the exact correlators of degree no more than 4.
Let us list some correlators as follows:
\begin{eqnarray*}
&&\langle {\textcolor{red}{\mathcal{K}_{2}}} \rangle
=\frac{\textcolor{red}{\tilde{\mathcal{N}}_r}\mathcal{N}_r}{\mu^2}
=\frac{\textcolor{red}{N_{1}}\mathcal{N}_r}{\mu^2}+\frac{\textcolor{green}{N_{2}}\textcolor{blue}{N_{3}}\cdots N_{r}\mathcal{N}_r}{\mu^2}
\nonumber\\
&&\includegraphics[height=2.25cm]{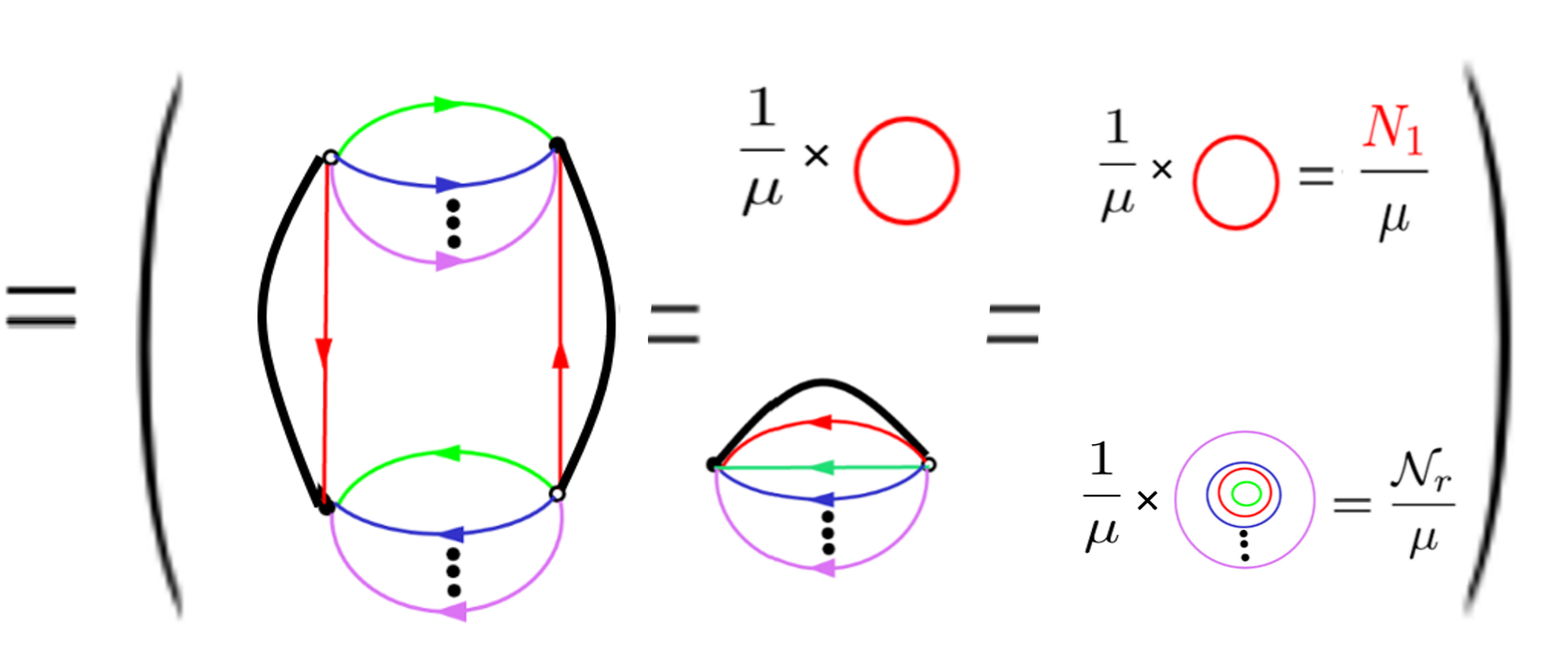}\includegraphics[height=2.25cm]{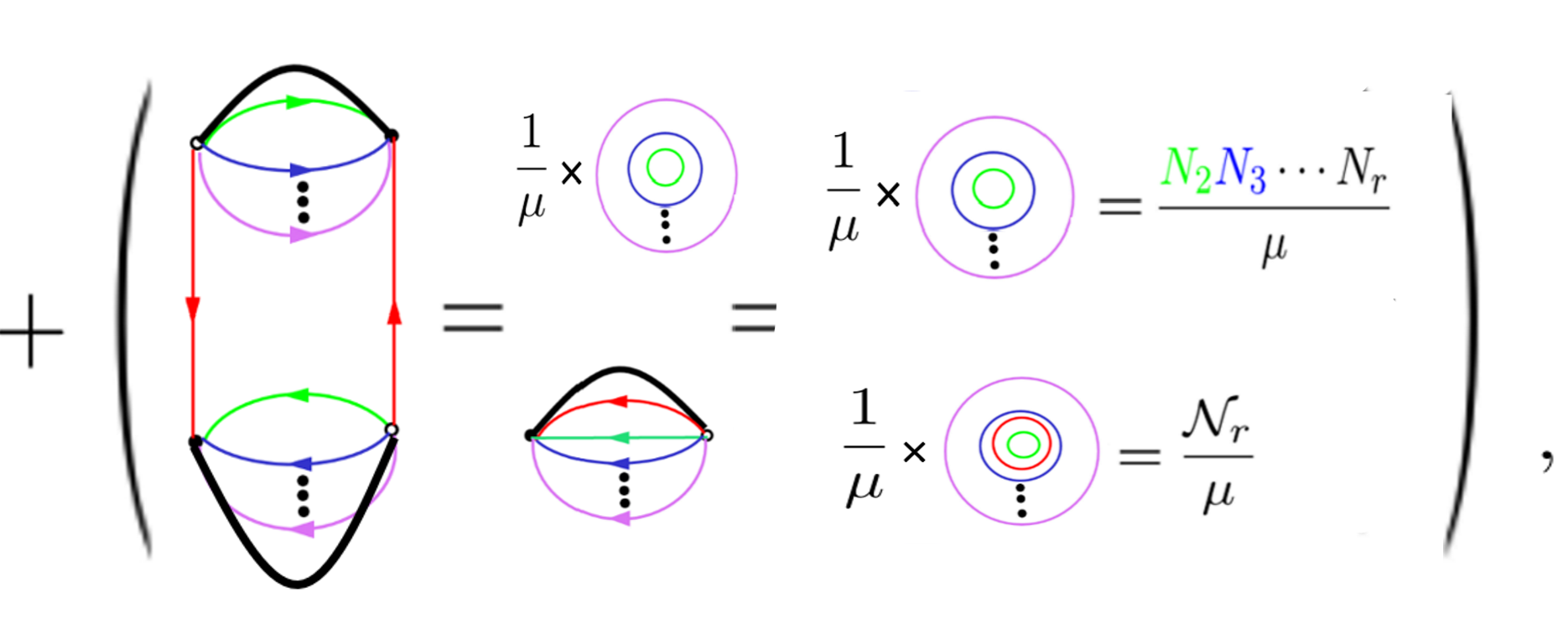}
\end{eqnarray*}
\begin{eqnarray*}
&&\langle {\textcolor{red}{\mathcal{K}_{1}\mathcal{K}_{2}}} \rangle
=\frac{\textcolor{red}{N_{1}}\mathcal{N}_r^2}{\mu^3}+\frac{\textcolor{green}{N_{2}}\textcolor{blue}{N_{3}}\cdots N_{r}\mathcal{N}_r^2}{\mu^3}
+2\frac{\textcolor{red}{N_{1}}\mathcal{N}_r}{\mu^3}+2\frac{\textcolor{green}{N_{2}}\textcolor{blue}{N_{3}}\cdots N_{r}\mathcal{N}_r}{\mu^3}
\nonumber\\
&&\includegraphics[height=3cm]{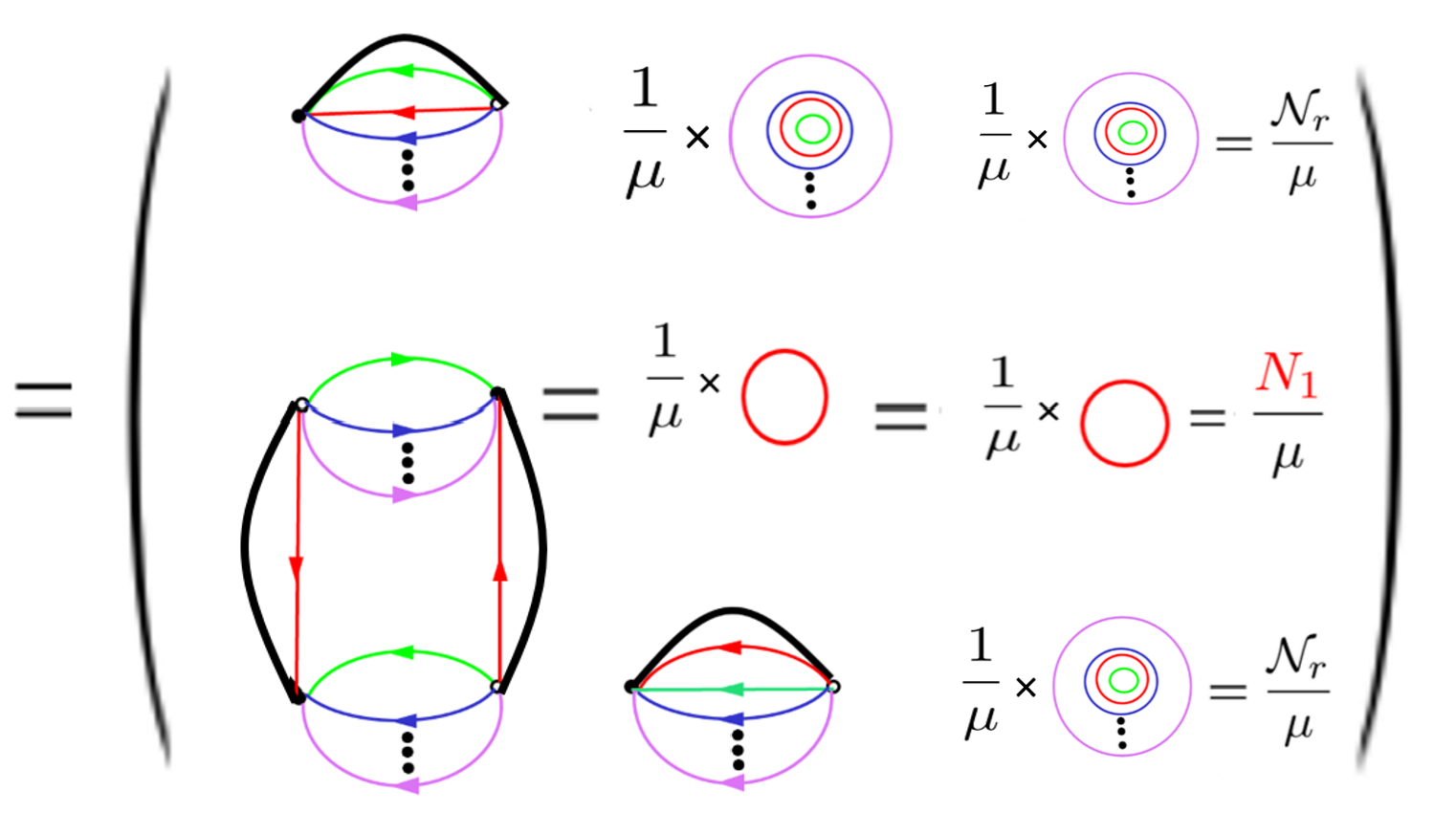}
\includegraphics[height=3cm]{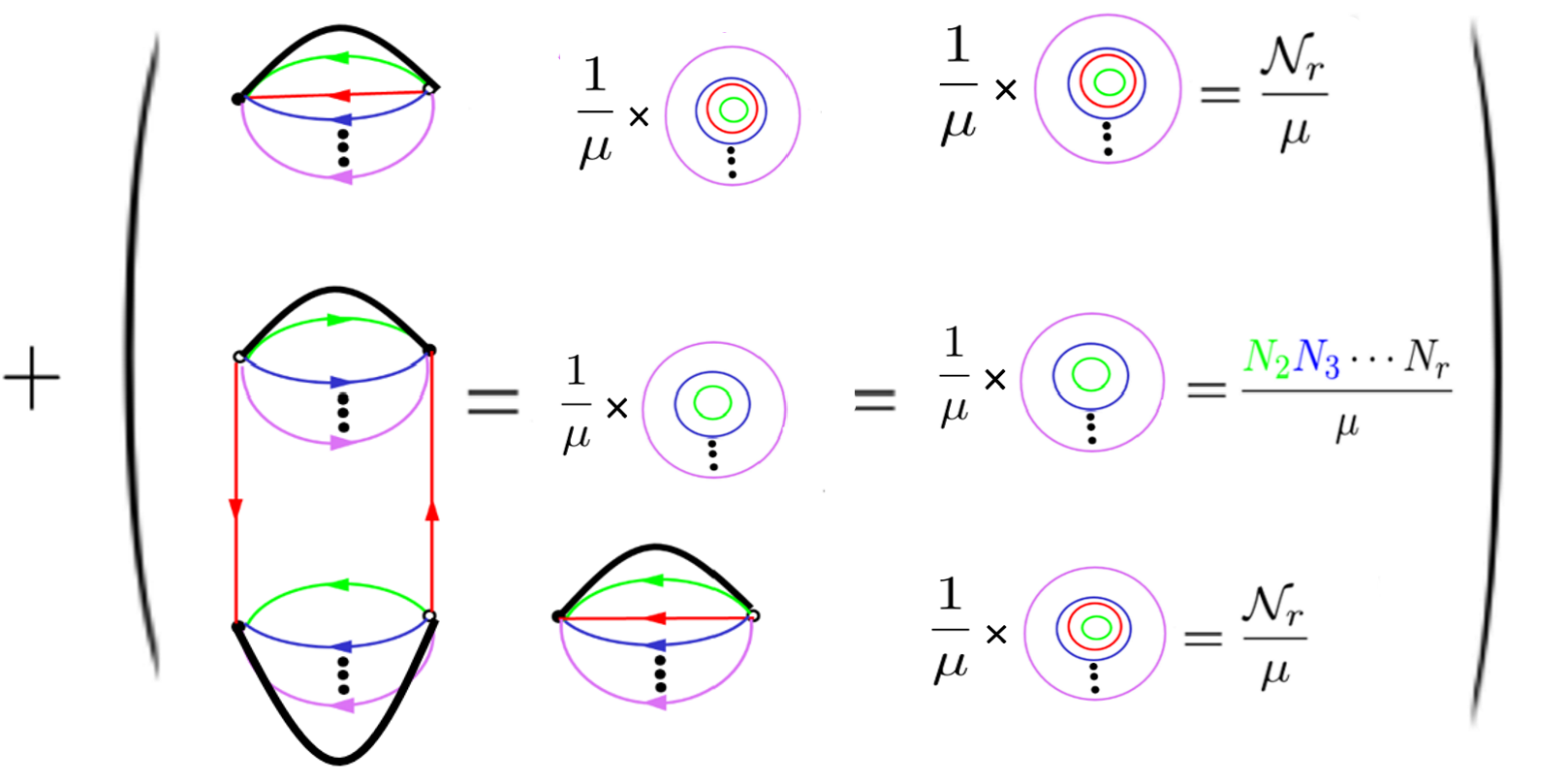}
\nonumber\\
&&\includegraphics[height=2.65cm]{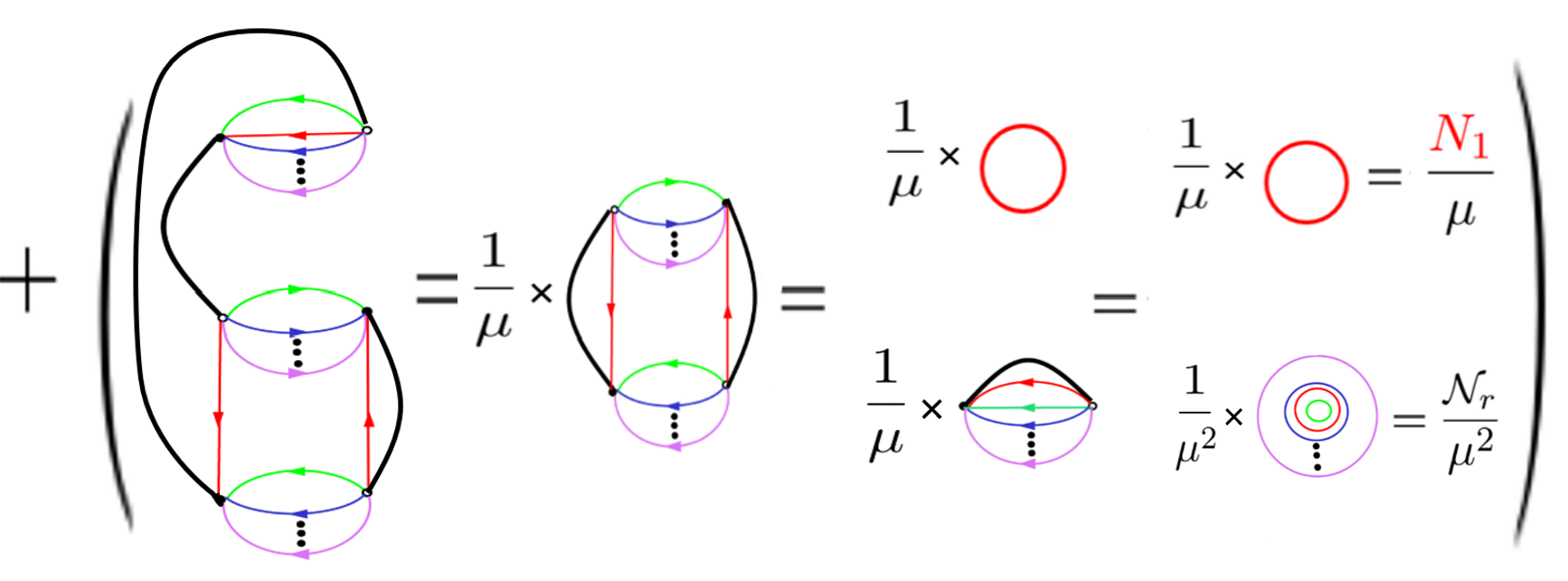}
\includegraphics[height=2.75cm]{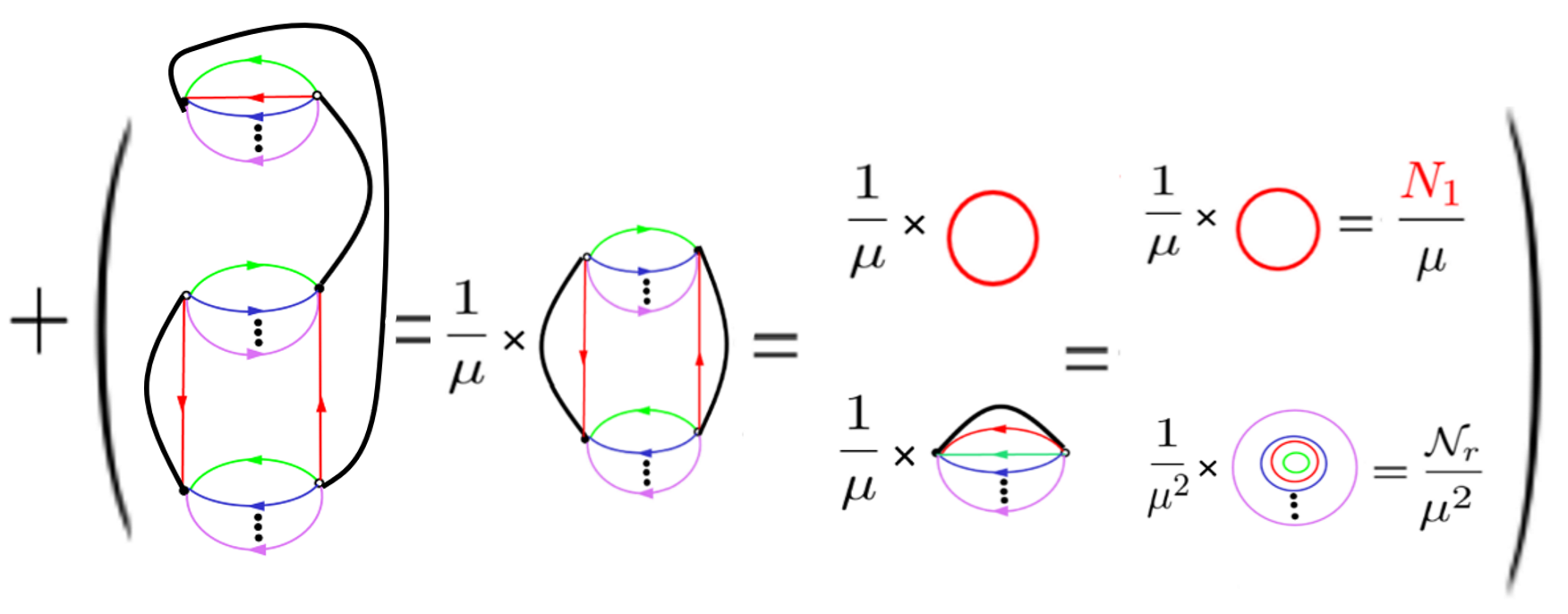}
\nonumber\\
&&\includegraphics[height=2.45cm]{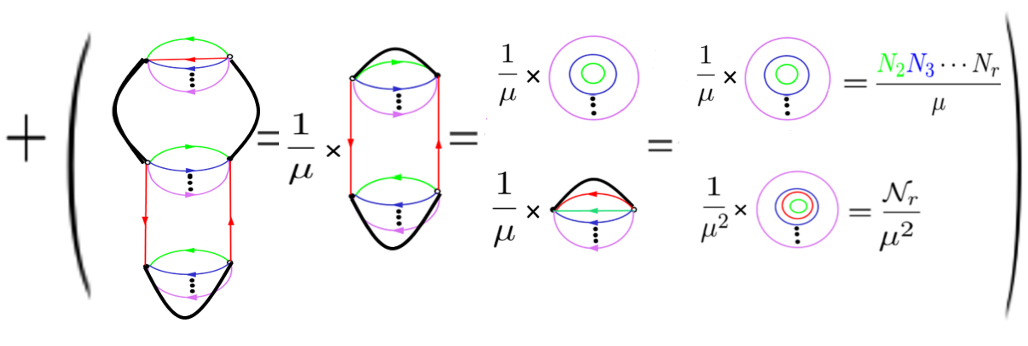}
\includegraphics[height=2.45cm]{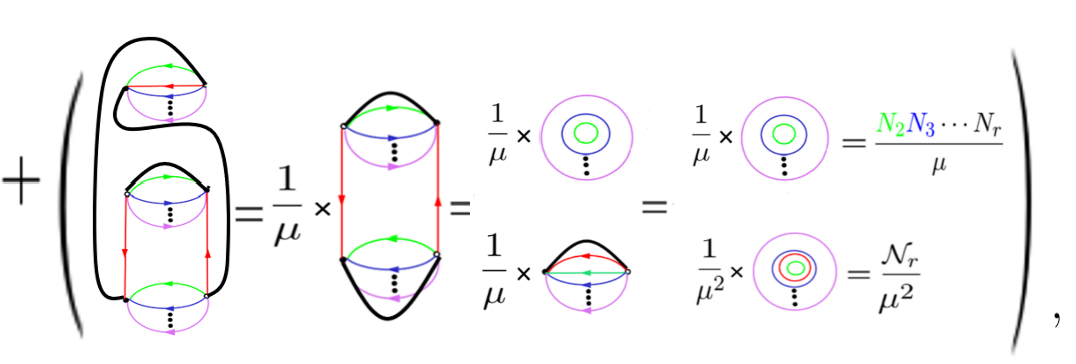}
\end{eqnarray*}
\begin{eqnarray}
&&\langle {\textcolor{red}{\mathcal{K}_{3}}} \rangle
=\frac{\textcolor{red}{N_1}^2\mathcal{N}_r}{\mu^3}
+2\frac{\textcolor{red}{N_1}\cdot\textcolor{green}{N_{2}}\textcolor{blue}{N_{3}}\cdots N_{r}\mathcal{N}_r}{\mu^3}
+\frac{(\textcolor{green}{N_{2}}\textcolor{blue}{N_{3}}\cdots N_{r})^2\mathcal{N}_r}{\mu^3}
+\frac{\mathcal{N}_r^2}{\mu^3}
+\frac{\mathcal{N}_r}{\mu^3}
\nonumber\\
&&\includegraphics[height=2.1cm]{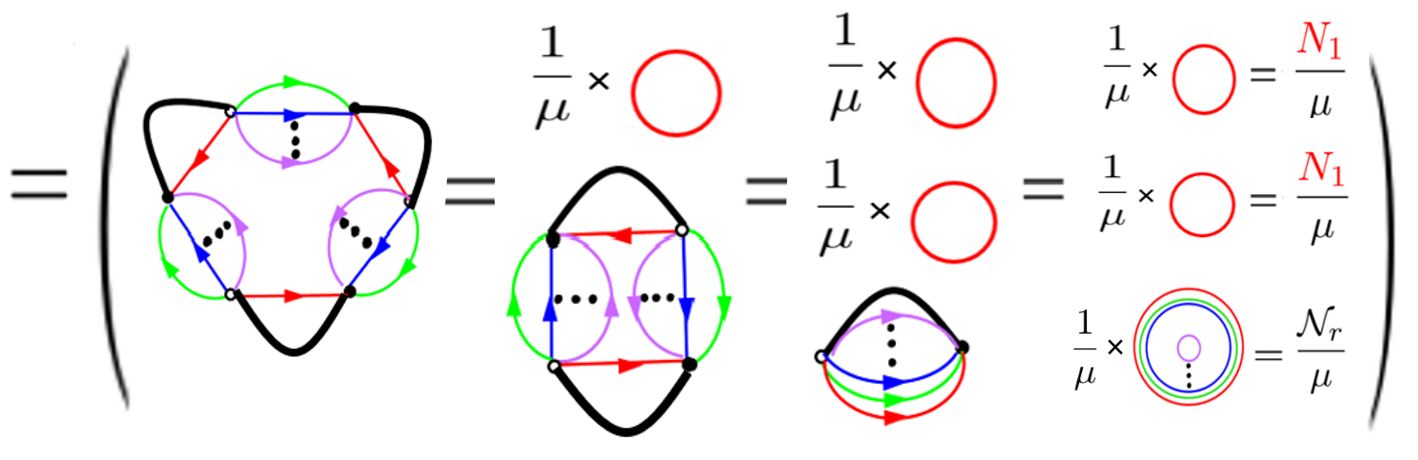}
\includegraphics[height=2.25cm]{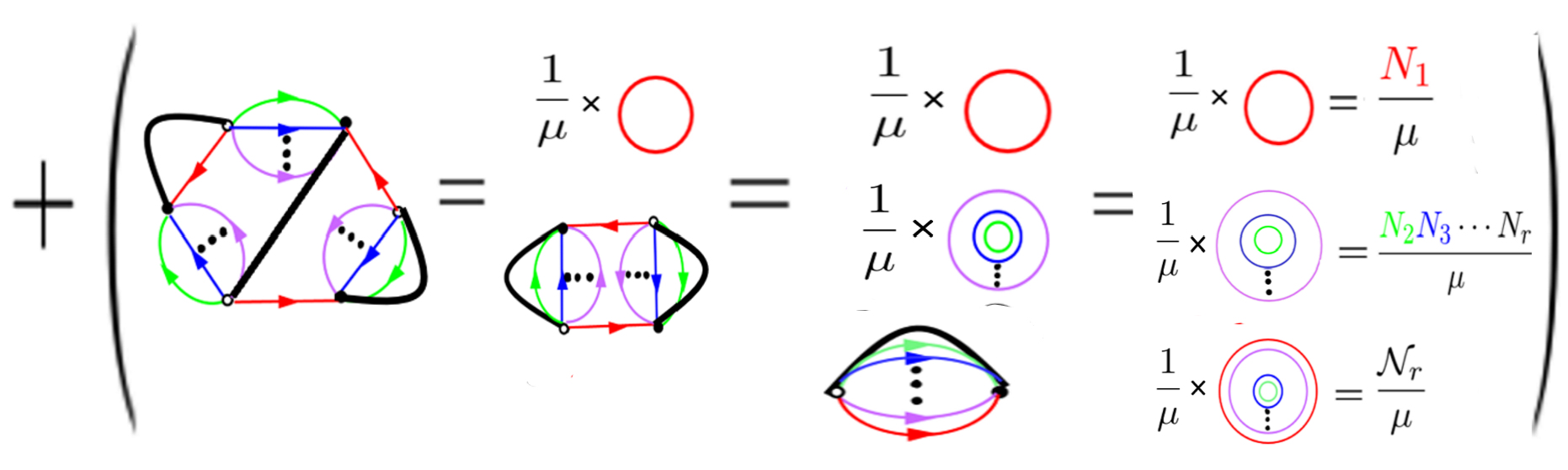}
\nonumber\\
&&\includegraphics[height=2.25cm]{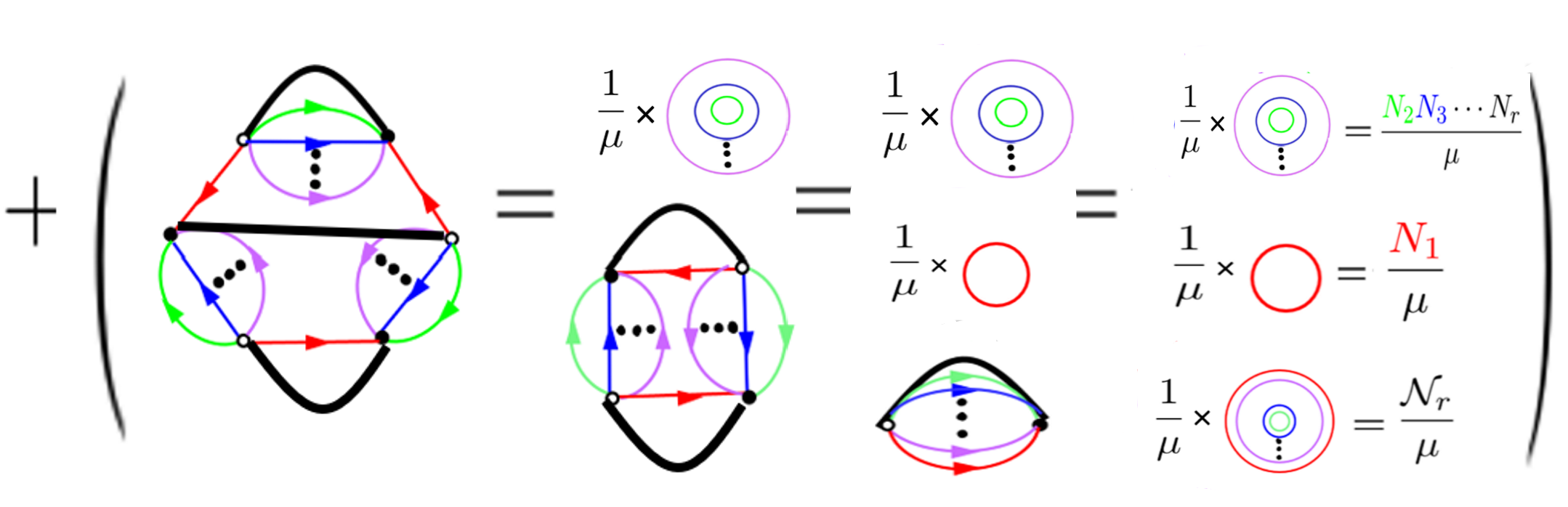}
\includegraphics[height=2.37cm]{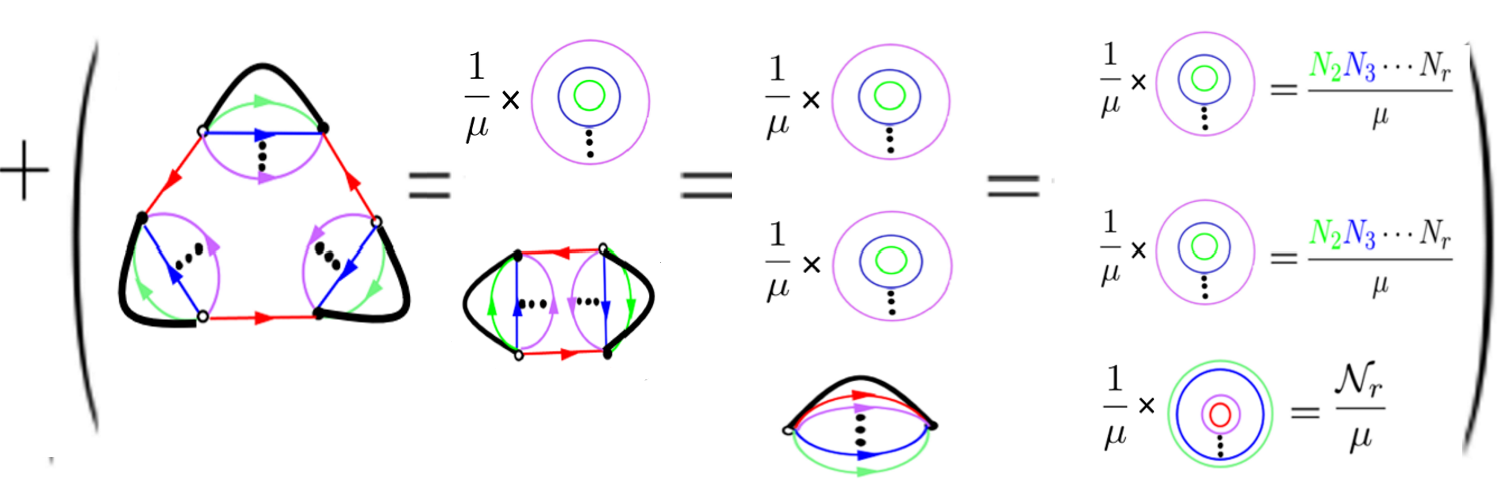}
\nonumber\\
&&\includegraphics[height=1.65cm,width=7cm]{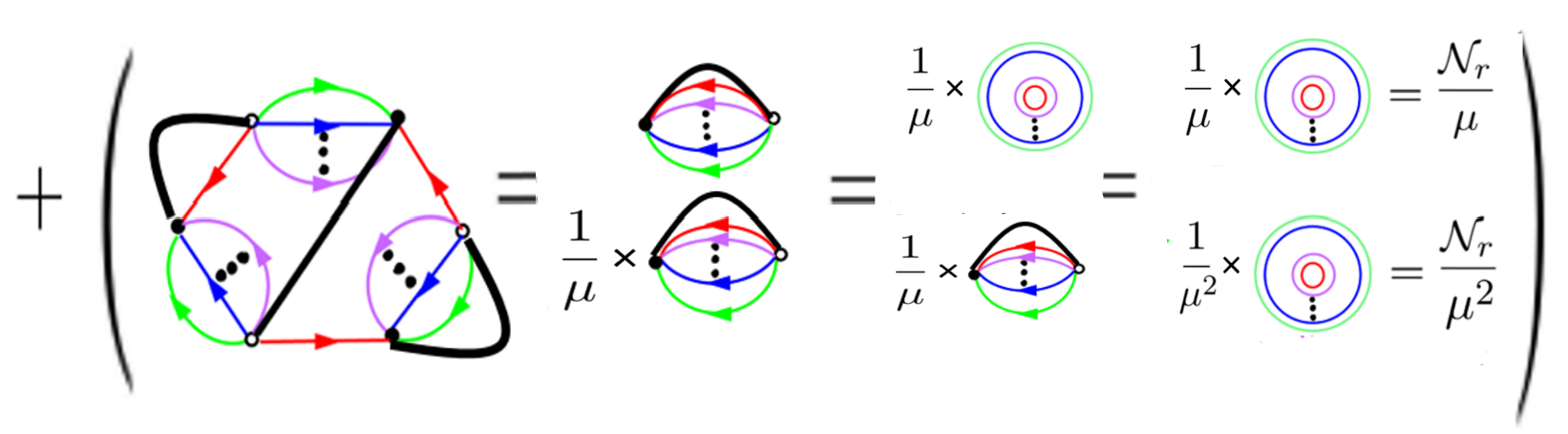}
\includegraphics[height=1.35cm,width=7cm]{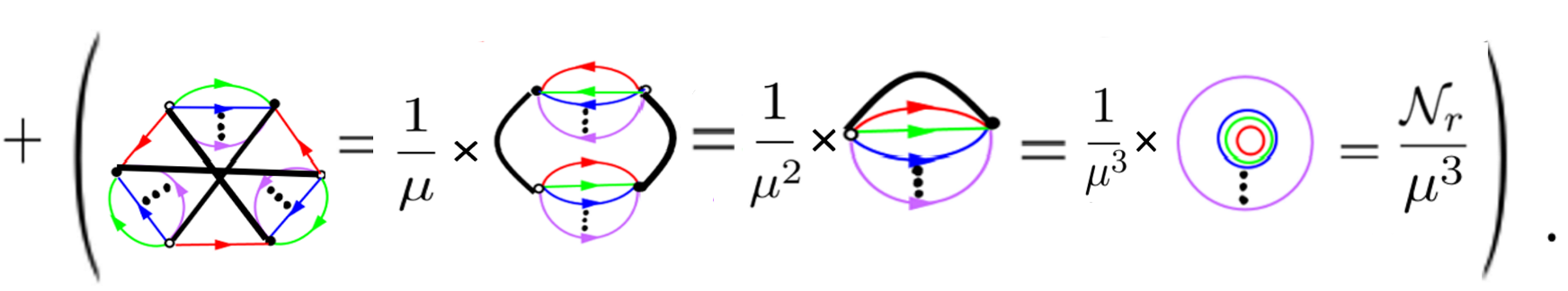}\nonumber\\
\end{eqnarray}

\subsection{Non-Gaussian red tensor model}

The correlators in the matrix models including supereigenvalue models have attracted considerable attention.
Much interest has also been attributed to the non-Gaussian cases \cite{Alexandrov04}-\cite{Cassia}.
But so far, no investigation has been made for the non-Gaussian tensor models.
The red tensor model is a simple rainbow tensor model. In the previous section, we have presented
its $W$-representation and the correlators.
Let us now focus on the non-Gaussian red tensor model
\begin{eqnarray}\label{NG-trivialtensor}
\textcolor{red}{Z_{NG}\{t,T\}}&=&
\int dA d\bar A \exp(-\frac{\mu}{p}\textcolor{red}{\mathcal{K}_{p}}-\mu\sum_{n=1}^{p-1} \textcolor{red}{T_{n}\mathcal{K}_{n}}
+\sum_{n=1}^{\infty}\textcolor{red}{t_n}\textcolor{red}{\mathcal{K}_{n}}).
\end{eqnarray}
In this model, the keystone operator is $\textcolor{red}{\mathcal{K}_2}$ (\ref{keystonetrivial}),
the tree and loop operators are the same with the case of the red tensor model.

From the deformation $\delta A=\dfrac{\partial \textcolor{red}
{\mathcal{K}_{m+1}}}{\partial \bar A}$ of the integration variable,
we derive the Virasoro constrains
\begin{eqnarray}\label{}
\hat{L}_m\textcolor{red}{Z_{NG}\{t,T\}}=0,
\end{eqnarray}
where
\begin{eqnarray}\label{NGRTV}
\hat{L}_m&=&\delta_{m,0}\mathcal{N}_r+(1-\delta_{m,0})\textcolor{red}{\tilde{\mathcal{N}_r}}
\textcolor{red}{\frac{\partial}{\partial t_{m}}}
+\sum_{b=1}^{m-1}\textcolor{red}{\frac{\partial}{\partial t_{b}}\frac{\partial}{\partial t_{m-b}}}\nonumber\\
&&-\mu\textcolor{red}{\frac{\partial}{\partial t_{m+p}}}
-\mu\sum_{n=1}^{p-1}n\textcolor{red}{T_n
\frac{\partial}{\partial t_{m+n}}}+\sum_{n=1}^{\infty}n
\textcolor{red}{t_n\frac{\partial}{\partial t_{m+n}}}.
\end{eqnarray}
The operators yield the Witt algebra (\ref{Valg}), but
the null 3-algebra (\ref{3Valg}) does not hold.

Let us take the deformation
$\delta A=\sum_{m=0}^{\infty}(m+p)\textcolor{red}{t_{m+p}}\dfrac{\partial \textcolor{red}
{\mathcal{K}_{m+1}}}{\partial \bar A}$ of the integration variable
in the integral (\ref{NG-trivialtensor}),
we obtain
\begin{eqnarray}\label{NG-trivialtensor-Constraints1}
(\mu\hat{D}_p-\hat{W}_{NG})\textcolor{red}{Z_{NG}\{t,T\}}=0,
\end{eqnarray}
where
\begin{eqnarray}\label{NG-trivial-D}
\hat{D}_p=\sum_{m=p}^{\infty}m\textcolor{red}{t_{m}\frac{\partial}{\partial t_m}},
\end{eqnarray}
and
\begin{eqnarray}\label{NG-trivial-w}
\hat{W}_{NG}&=&\sum_{m=1}^{\infty}\sum_{b=1}^{m-1}(m+p)
\textcolor{red}{t_{m+p}\frac{\partial}{\partial t_{b}}
\frac{\partial}{\partial t_{m-b}}}
+p\mathcal{N}_r\textcolor{red}{t_{p}}+\sum_{n=1}^{\infty}\sum_{m=0}^{\infty}n(m+p)
\textcolor{red}{t_nt_{m+p}\frac{\partial}{\partial t_{m+n}}}\nonumber\\
&&
+\sum_{m=1}^{\infty}(m+p)\textcolor{red}{\tilde{\mathcal{N}_r}}
\textcolor{red}{{t_{m+p}
\frac{\partial}{\partial t_{m}}}}
-\mu\sum_{n=1}^{p-1}\sum_{m=0}^{\infty}n(m+p)\textcolor{red}{T_nt_{m+p}
\frac{\partial}{\partial t_{m+n}}}.
\end{eqnarray}

On the other hand, there are the additional constraints for the partition function (\ref{NG-trivialtensor})
\begin{eqnarray}\label{NG-trivialtensor-Constraints2}
(\mu\textcolor{red}{\frac{\partial}{\partial t_m}}+\textcolor{red}{\frac{\partial}{\partial T_m}})\textcolor{red}{Z_{NG}\{t,T\}},\quad m=1,\cdots,p-1.
\end{eqnarray}

Combining (\ref{NG-trivialtensor-Constraints2}) and (\ref{NG-trivialtensor-Constraints1}), we have
\begin{eqnarray}\label{NG-Constraints DW}
(\mu \hat{\mathcal{D}}-\hat{\mathcal{W}})\textcolor{red}{Z_{NG}\{t,T\}}=0,
\end{eqnarray}
where $\hat{\mathcal{D}}=\dsum_{m=0}^{\infty}m\textcolor{red}{t_{m}\frac{\partial}{\partial t_m}}$
and $\hat{\mathcal{W}}=\hat{W}_{NG}-\dsum_{m=0}^{p-1}m\textcolor{red}{t_{m}\frac{\partial}{\partial T_m}}$.

By expanding the exponential term in (\ref{NG-trivialtensor}), we rewrite the partition function as
\begin{eqnarray}
\textcolor{red}{Z_{NG}\{t,T\}}&=&\sum_{s=0}^{\infty}\textcolor{red}{Z_{NG}^{(s)}\{t,T\}}
\nonumber\\
&=&\textcolor{red}{Z_{NG}\{T\}}\cdot(1+\sum_{n_1=1}^{\infty}\textcolor{red}{t_{n_1}}
\langle\textcolor{red}{\mathcal{K}_{n_1}}\rangle_{NG}
+\sum_{n_1,n_2=1}^{\infty}\textcolor{red}{t_{n_1}t_{n_2}}\langle\textcolor{red}
{\mathcal{K}_{n_1}\mathcal{K}_{n_2}}\rangle_{NG}+\cdots),
\end{eqnarray}
where
\begin{eqnarray}
\textcolor{red}{Z_{NG}^{(s)}\{t,T\}}&=&\textcolor{red}{Z_{NG}\{T\}}\cdot\sum_{l=0}^{\infty}\sum_{n_1+\cdots +n_l=s}\frac{1}{l!} \textcolor{red}{t_{n_1}t_{n_2}\cdots t_{n_l}}\langle\textcolor{red}{\mathcal{K}_{{n_1}}\mathcal{K}_{{n_2}}\cdots\mathcal{K}_{{n_l}}}\rangle_{NG},
\end{eqnarray}

\begin{eqnarray}
\textcolor{red}{Z_{NG}\{T\}}&=&\int dA d\bar A\exp[-\frac{\mu}{p}\textcolor{red}{\mathcal{K}_{p}}-\mu\sum_{n=1}^{p-1} \textcolor{red}{T_{n}\mathcal{K}_{n}}],
\end{eqnarray}

the correlators are defined by
\begin{eqnarray}
%C_{n_1\cdots n_l}(T)&=&
\langle\textcolor{red}{\mathcal{K}_{{n_1}}\mathcal{K}_{{n_2}}\cdots\mathcal{K}_{{n_l}}}\rangle_{NG}
=\frac{\int dA d\bar A \textcolor{red}{\mathcal{K}_{{n_1}}\mathcal{K}_{{n_2}}\cdots\mathcal{K}_{{n_l}}} \exp(-\dfrac{\mu}{p}\textcolor{red}{\mathcal{K}_{p}}
-\mu\dsum_{n=1}^{p-1} \textcolor{red}{T_{n}\mathcal{K}_{n}})}{\int dA d\bar A \exp(-\dfrac{\mu}{p}\textcolor{red}{\mathcal{K}_{p}}
-\mu\dsum_{n=1}^{p-1} \textcolor{red}{T_{n}\mathcal{K}_{n}})}.
\end{eqnarray}

The operators $\hat{\mathcal{D}}$ and $\hat{\mathcal{W}}$ acting on $\textcolor{red}{Z_{NG}^{(s)}\{t,T\}}$
show that $\hat{\mathcal{D}}$ is the operator preserving the grading.  However, $\hat{\mathcal{W}}$ is not the desired
operator increasing the grading, since it does not satisfy the similar relation (\ref{increa}).
Unlike the case of (\ref{exp}), the partition function can not be realized by acting on elementary function
with exponents of the operator $\hat{\mathcal{W}}$.

Note that $\hat{\mathcal{D}}\textcolor{red}{Z_{NG}\{T\}}=0$. From (\ref{NG-Constraints DW}),
it is not difficult to obtain that
\begin{eqnarray}\label{NG-trivialteensor-level}
\sum_{s=1}^{\infty}\textcolor{red}{Z_{NG}^{(s)}\{t,T\}}&=&(\mu \hat{\mathcal{D}}-\hat{\mathcal{W}})^{-1}\hat{\mathcal{W}}\textcolor{red}{Z_{NG}\{T\}}
\nonumber\\&=&\sum_{k=1}^{\infty}\mu^{-k}( \hat{\mathcal{D}}^{-1}\hat{\mathcal{W}})^{k}\textcolor{red}{Z_{NG}\{T\}}.
\end{eqnarray}

Thus the partition function (\ref{NG-trivialtensor}) can be expressed as
\begin{eqnarray}\label{NG-trivialtensor2}
\textcolor{red}{Z_{NG}\{t,T\}}=\sum_{k=0}^{\infty}\mu^{-k}( \hat{\mathcal{D}}^{-1}\hat{\mathcal{W}})^{k}
\textcolor{red}{Z_{NG}\{T\}}.
\end{eqnarray}
We observe that (\ref{NG-trivialtensor2}) is similar with the case of non-Gaussian matrix model \cite{Cassia}.
It shows that the dual expression for the non-Gaussian red tensor model (\ref{NG-trivialtensor})
through differentiation can also be formulated.

Since the usual $W$-representation of (\ref{NG-trivialtensor}) fails,
we can not present the compact expression of correlators here.
In principle, we can give the correlators from (\ref{NG-trivialtensor2}). Let us list some correlators as follows:
\begin{eqnarray}\label{NG-trivialtensor-corre}
\langle\textcolor{red}{\mathcal{K}_{{i}}}\rangle_{NG}&=&
-\frac{1}{\mu}\frac{1}{\textcolor{red}{Z_{NG}\{T\}}}\textcolor{red}{\frac{\partial}{\partial T_i}Z_{NG}\{T\}},\ \ (i=1,\cdots,p-1), \nonumber\\
\langle\textcolor{red}{\mathcal{K}_{{p}}}\rangle_{NG}&=&\frac{\mathcal{N}_r}{\mu}+
\frac{1}{\mu} \sum_{i=1}^{p-1}\frac{i\textcolor{red}{T_i}}{\textcolor{red}{Z_{NG}\{T\}}}\textcolor{red}{\frac{\partial}{\partial T_i}Z_{NG}\{T\}},
\nonumber\\
\langle\textcolor{red}{\mathcal{K}_{{p+1}}}\rangle_{NG}&=&
-\frac{\textcolor{red}{\tilde{\mathcal{N}_r}}}{\mu^2}\textcolor{red}{\frac{\partial}{\partial T_1}Z_{NG}\{T\}}
+\frac{1}{\mu} \sum_{i=1}^{p-2}\frac{i\textcolor{red}{T_i}}{\textcolor{red}{Z_{NG}\{T\}}}\textcolor{red}{\frac{\partial}{\partial T_{i+1}}Z_{NG}\{T\}}
\nonumber\\
&&
-\frac{\mathcal{N}_r}{\mu}(p-1)\textcolor{red}{T_{p-1}}
-(p-1)\textcolor{red}{T_{p-1}} \sum_{i=1}^{p-1}\frac{i\textcolor{red}{T_i}}{\textcolor{red}{Z_{NG}\{T\}}}\textcolor{red}{\frac{\partial}{\partial T_{i}}Z_{NG}\{T\}},
\nonumber\\
\langle\textcolor{red}{\mathcal{K}_{i_{1}}\mathcal{K}_{i_{2}}}\rangle_{NG}
&=&\frac{1}{\mu^{2}}\frac{1}{\textcolor{red}{Z_{NG}\{T\}} }\textcolor{red}{\frac{\partial^2}{\partial T_{i_1}\partial T_{i_2}}Z_{NG}\{T\}},
(i_1, i_2=1,\cdots,p-1),
\nonumber\\
\langle\textcolor{red}{\mathcal{K}_{{p}}\mathcal{K}_{{q}}}\rangle_{NG}
&=&
-\frac{1}{\mu^2}\sum_{i=1}^{p-1}\frac{i\textcolor{red}{T_i}}{\textcolor{red}{Z_{NG}\{T\}}}
\textcolor{red}{\frac{\partial^2}{\partial T_i\partial T_q}Z_{NG}\{T\}}\nonumber\\
&-&\frac{\mathcal{N}_r+q}{\mu^2}\textcolor{red}{\frac{\partial}{\partial T_q}Z_{NG}\{T\}}, \ \ (q=1,\cdots,p-1).
\end{eqnarray}

%%%%%%%%%%%%%%%%%%%%%%%%%%%%%%%%%%%%%%%%%%%%%%%%%%%%%%%%%%%
\section{Conclusions}
%%%%%%%%%%%%%%%%%%%%%%%%%%%%%%%%%%%%%%%%%%%%%%%%%%%%%%%%%%%
$W$-representation is important for the understanding of matrix
model, since it provides a dual formula for partition function through
differentiation.
We have investigated the $W$-representation of the rainbow tensor model
with the rank $r$ complex tensor in this paper.
By the given deformation of the integration variable in the integral, we derived the desired
operators $\hat D_r$ and $\hat W_r$ which preserve and increase the grading, respectively.
It was shown that the rainbow tensor model can be realized
by acting on elementary function with exponent of the operator $\hat W_r$.
In terms of the operators preserving and increasing the grading,
we can construct the Virasoro constraints for the rainbow tensor model,
where the constraint operators obey the Witt algebra and null 3-algebra.
An interesting aspect of these Virasoro constraints is that
the compact expression of correlators can be derived from them.
As examples, we have applied above results to analyze the red tensor model,
Aristotelian tensor model and $r=4$ rainbow tensor model in detail and presented the
corresponding correlators in these models.

We have also considered the non-Gaussian red tensor model and presented the Virasoro constraints,
where the constraint operators obey the Witt algebra, however the null 3-algebra  does not hold.
We showed that the partition function can be expressed as the
infinite sum of the operators $(\hat{\mathcal{D}}^{-1}\hat{\mathcal{W}})^{k}$ acting on the given
function. Namely, a dual form for partition function through differentiation can be formulated.
Since $\hat{\mathcal{W}}$ is not the desired operator increasing the grading,
it causes the usual $W$-representation of the non-Gaussian red tensor model to fail here.
For this reason,  the dual expression (\ref{NG-trivialtensor2}) through differentiation
can be regarded as the generalized $W$-representation.
In terms of the operators $\hat{\mathcal{D}}$ and $\hat{\mathcal{W}}$,
we can not construct the Virasoro constraints such that the constraint operators obey the Witt algebra
and null 3-algebra.
It should be noted that we can calculate the correlators from (\ref{NG-trivialtensor2}).
However, the compact expression of correlators can not be derived. We have presented
some correlators. How to represent these correlators graphically still deserves further study.
Furthermore, further study should be done
to investigate the $W$-representations of the non-Gaussian and fermionic tensor models.

\section *{Acknowledgments}

This work is supported by the National Natural Science Foundation
of China (Nos. 11875194 and 11871350).

%%%%%%%%%%%%%%%%%%%%%%%%%%%%%%%%%%%%%%%%%%%%%%%%%%%%%%%%%%%%%%%%%%%%%%%%%%%%%%%

\end{document}